\documentclass[superscriptaddress,preprintnumbers,preprint,showpacs,amsmath,amssymb]{revtex4-1}
\usepackage{amsmath}
\usepackage{psfrag}
\usepackage{graphicx}
\usepackage{subfigure}
\usepackage{array}
\usepackage{natbib} 
\usepackage{xcolor}
\def\be{\begin{equation}}       \def\ee{\end{equation}}
\def\bea{\begin{eqnarray}}      \def\eea{\end{eqnarray}}

\begin{document}
\title{Asymptotic entanglement sudden death in two atoms with dipole-dipole and Ising interactions coupled to a radiation field at non-zero detuning}
\author{Gehad Sadiek\footnote{Corresponding author: gsadiek@sharjah.ac.ae}}
\affiliation{Department of Applied Physics and Astronomy, University of Sharjah, Sharjah 27272, UAE}
\affiliation{Department of Physics, Ain Shams University, Cairo 11566, Egypt}
\author{Wiam Al-Drees}
\affiliation{Department of Physics, Imam Muhammad Ibn Saud Islamic University, Riyadh 11432, Saudi Arabia}
\author{Salwa Shaglel}
\affiliation{Department of Physics, University of Siegen, Siegen 57068, Germany}
\author{Hala Elhag}
\affiliation{Department of Physics, University of Hamburg, Hamburg 22589, Germany}
\begin{abstract}
We investigate the time evolution and asymptotic behavior of a system of two two-level atoms (qubits) interacting off-resonance with a single mode radiation field. The two atoms are coupled to each other through dipole-dipole as well as Ising interactions. An exact analytic solution for the system dynamics that spans the entire phase space is provided. We focus on initial states that cause the system to evolve to entanglement sudden death (ESD) between the two atoms.
We find that combining the Ising and dipole-dipole interactions is very powerful in controlling the entanglement dynamics and ESD compared with either one of them separately. Their effects on eliminating ESD may add up constructively or destructively depending on the type of Ising interaction (Ferromagnetic or anti-Ferromagnetic), the detuning parameter value, and the initial state of the system. 
The asymptotic behavior of the ESD is found to depend substantially on the initial state of the system, where ESD can be entirely eliminated by tuning the system parameters except in the case of an initial correlated Bell state. Interestingly, the entanglement, atomic population and quantum correlation between the two atoms and the field synchronize and reach asymptotically quasi-steady dynamic states. Each one of them ends up as a continuous irregular oscillation, where the collapse periods vanish, with a limited amplitude and an approximately constant mean value that depend on the initial state and the system parameters choice. This indicates an asymptotic continuous exchange of energy (and strong quantum correlation) between the atoms and the field takes place, accompanied by diminished ESD for these chosen setups of the system. 
This system can be realized in spin states of quantum dots or Rydberg atoms in optical cavities, and superconducting or hybrid qubits in linear resonators.
\end{abstract}
\maketitle
\section{Introduction}
\label{Introduction}

Studying the quantum phenomena in systems of atoms coupled to radiation fields has been in the center of interest in physics since the Jaynes-Cummings model was introduced in 1963 \cite{Jaynes-cummings:1963}. This interest was boosted by the development of new quantum structures that are considered very promising for serving as the building units in quantum information processing (QIP) systems \cite{Nielsen-Chuang:2010}, while at the same time are highly interacting with and controllable by radiation fields, which include artificial atomic systems such as semiconducting quantum dots and superconducting circuits in addition to the customized atomic systems such as Rydberg atoms and trapped atoms, ions and molecules \cite{Buluta:2011, Xiang:2013, Lodahl:2015, Wendin:2017}. Interaction between natural regular atoms, in cavity quantum electrodynamics (CQED), used to be ignored as a result of their considerably small magnitude. However, the newly developed quantum systems are characterized by strong interaction with the same type of system or even other types in hybrid quantum composite systems.
Embedding superconducting qubits in a superconducting microwave resonator was a huge step toward utilizing these systems in QIP \cite{Yang:2003,Yang:2004,Blais:2004} and establishing the new field of circuit quantum electrodynamics (cQED), where close and distant superconducting qubits can be coupled through local interactions or microwave photons \cite{Yamamoto:2003, Majer:2005, Steffen:2006, Van:2007, Niskanen:2007}. Also, other systems can be embedded in superconducting resonators, through hybrid circuit structures, such as spins in quantum dots and solid state impurities \cite{Xiang:2013, Wendin:2017}. Similar arrangements were implemented for coupled Rydberg atoms in CQED \cite{Osnaghi:2001,Gywat:2006,Guerlin:2010,Donaire:2017}.
Laser-trapped circular Rydberg atoms were used for analogue quantum simulation of spin arrays \cite{Nguyen:2018}, where the strong interaction between the atoms is used to simulate an XXZ spin chain Hamiltonian, which amounts for dipole-dipole and Ising interactions between the spins. It was suggested that such a scheme can be also utilized in hybrid structures in cQED, where Rydberg atoms can be integrated into superconducting circuits.
Spin-spin interaction simulation and modeling in other systems such as optical lattices \cite{Sorensen:1999}, trapped ions \cite{Porras:2004}, coupled molecules \cite{Cusati:2006,Napoli:2006} and microcavities \cite{Hartmann:2007} were performed.
These important developments led to a growing interest in studying decoherence and entanglement dynamics, and population inversion in systems of atoms (qubits), with either dipole-dipole or Ising interaction, or both, in the presence of radiation fields, keeping an eye on their QIP implementations. 
Coupled spin (qubit) systems in absence of radiation fields but in presence of magnetic fields and different types of environments have been studied intensively \cite{Khlebnikov:2002, Huang:2006, Abliz:2006, Tsomokos:2007, Buric:2008, Dubi:2009, Sadiek:2010, Xu:2011, Sadiek:2013, Duan:2013, Wu:2014, Sadiek:2016}. 

One of the main obstacles toward realizing reliable quantum computing systems in general, and particularly using these newly customized quantum systems, is entanglement sudden death (ESD). ESD is observed when the entanglement loss in the system takes place very rapidly leading to a disentangled state. It was first identified, discussed and named by Yu and Eberly \cite{Yu:2004,Yu:2008}, when they studied the entanglement between two uncoupled, but initially entangled, atoms in two separate cavities. They also showed that the same phenomenon may take place under the effect of a noisy classical environment on two uncoupled atoms \cite{Yu:2006}.
Several other works have studied ESD in systems of two non-interacting atoms in remote cavities as well \cite{Yonac:2006, Sainz:2006, Yonac:2007, Sainz:2007}. The effect of an out of resonance radiation field on systems of identical and non-identical, non-interacting, atoms was studied, where it was shown that the non-zero detuning can be an advantage for preserving entanglement \cite{Chan:2009, Tavassoly:2018}. The Bell inequality was tested using a system of two uncoupled qubits interacting with a radiation field in an optical cavity \cite{Bai:2017}.
Although systems of interacting atoms have been considered as well, but they were only at resonance with the field to avoid the mathematical difficulty caused by the off-resonance condition \cite{Liu:2006,Ficek:2006, Deng:2016, Li:2009}. Particularly, ESD was studied in a system of two coupled identical atoms interacting at resonance with a double mode radiation field, where the effects of the coupling as well as the initial state of the system on the system dynamics were investigated \cite{Zhang:2007}. Entanglement and purity in a system of two interaction atoms coupled to a radiation field at resonance were investigated \cite{Torres:2010}, where the effect of the interplay between the atom-atom and the atom-field field couplings on the system was studied thoroughly.
A system of two coupled atoms interacting off-resonance with a radiation field was studied \cite{Gywat:2006}, where the system was represented by XXZ model, considering dipole-dipole and Ising interactions at the same time. They provided an analytic perturbative solution for the system dynamics assuming a weak atom-atom coupling. They proposed an experimental realization for the system using spins states in quantum dots in CQED and superconducting qubits in cQED. Another work considered two identical atoms that are coupled to each other through dipole-dipole and Ising interactions while coupled to a radiation field at resonance, where an analytic solution was provided \cite{Sanchez:2016}. To study the entanglement in the system numerical calculations were performed.
Recently, a system of two coupled qubits interacting with a common environment was investigated, where schemes to avoid ESD using Local unitary operations were provided \cite{ Chathavalappil:2019}. Very recently, the entanglement dynamics of a pair of well-separated Rydberg pairs driven by a common laser field while interacting via both intra-pair and inter-pair van der Waals potentials was investigated \cite{Zhang:2020}. It showed in-phase (anti-phase) beating dynamics that depends on the inter-pair potentials and the field detuning.

As can be noticed, in the previous works, the dynamics of the system of atoms coupled to a radiation field were studied intensively but under certain restrictions, due to the mathematical formidability of the problem. It was considered either uncoupled atoms interacting off-resonance with the field or coupled atoms (with a dipole, Ising, or both interactions) at resonance with the field. In a recent work \cite{Sadiek:2019}, we studied a system of two coupled atoms (qubits) with dipole-dipole interaction in the presence of an off-resonance radiation field. We presented an analytic solution for the time evolution of the system and showed how the combined effect of the dipole coupling and the non-zero detuning can be utilized to control the ESD in the system.

In this paper, we study a system of two coupled atoms interacting off-resonance with a single-mode radiation field. We provide an exact analytic study of the system dynamics while considering simultaneously the effect of the two types of interactions, dipole-dipole and Ising, which can be modeled as a Heisenberg spin 1/2 XXZ interaction. This system is important for its own sake as a model of two coupled spins interacting with a bosonic bath, as well as for its impact on the field of cavity (circuit) QED and its implementations in QIP. It can be realized in electron spin states in quantum dots or Rydberg atoms in optical cavities as well as in superconducting or hybrid qubits in linear resonators. This work represents a crucial completion of our previous one that is vitally needed to provide a full understanding of the system dynamics and asymptotic behavior.
We emphasize, using our results, the significant impact of the Ising interaction, on its own or when combined with the dipole one, on the system dynamics. Particularly, we show how the type of interaction (Ferromagnetic vs. anti-Ferromagnetic) may lead to considerably different effects on the system dynamics, especially the ESD, depending on the initial state and the detuning value.
Moreover, we investigate the asymptotic behavior of the system dynamics and particularly the ESD, under the interplay of the different system parameters, which have not been addressed before in the literature or in our previous work. Most of the previous works focused on the early dynamics of the system and the treatment of the ESD at that stage.
We show that the system reaches asymptotically a quasi-steady dynamic state, where the ESD can be eliminated by utilizing the interplay between the Ising and the dipole-dipole interactions as well as the detuning parameter for all initial states except the maximally entangled correlated Bell state. Also, we demonstrate how the entanglement, population inversion and quantum correlation between the two atoms ensemble and the radiation field synchronize asymptotically with remarkable profiles, where the collapse periods disappear indicating a continuous exchange of energy (and quantum correlation) between the atoms and the field, accompanied by diminished ESD for chosen setups of the system.

This paper is organized as follows. In Sec. 2, we discuss our model and present the exact analytic solution. In sec. 3, we study the dynamics of entanglement and atomic population inversion, starting from different initial states. The quantum correlation between the two atoms ensemble and the radiation field is presented in sec. 4. We conclude in Sec. 5.
\section{The model and its exact solution}
\label{The Model}
We consider a system of two identical atoms (qubits), each one of them is characterized by two levels: ground $\left|g_{i}\right\rangle$ and excited $\left|e_{i}\right\rangle$, where $i=1,2$ corresponding to the first and second atoms respectively. The two atoms are coupled to each other through dipole-dipole and Ising interactions, which are modeled as an $XXZ$ exchange interactions between two spin-1/2 particles with $XX$ coupling strength $\lambda_2$ and $Z$ coupling strength $J$. They are coupled to the same single-mode quantized radiation field with the same coupling constant $\lambda_1$. The Hamiltonian of the system is given by 
\begin{eqnarray}
 \hat{H} = \Omega \; \hat{a}^\dagger \hat{a} + \frac{\omega_{\circ}}{2} \sum_{i=1,2} \hat{\sigma}^{(i)}_{z} + \lambda_1 \sum_{i=1,2} (\hat{a}\hat{\sigma}^{(i)}_{+}+\hat{a}^\dagger \hat{\sigma}^{(i)}_{-}) + \lambda_2 \; (\hat{\sigma}^{(1)}_{-}\hat{\sigma}^{(2)}_{+}+\hat{\sigma}^{(1)}_{+}\hat{\sigma}^{(2)}_{-})  + J \sigma^{(1)}_{z} \sigma^{(2)}_{z}.
\label{eq:H}
\end{eqnarray}
The first and second terms in the Hamiltonian represent the free quantized radiation field and the non-interacting two atoms respectively. The third, fourth, and fifth terms represent the atom-field, dipole-dipole, and Ising interactions respectively. $\Omega$ and $\omega_{\circ}$ are the frequencies of the single-mode radiation field and the atomic system transition respectively, $\hat{a}^\dagger$ and $\hat{a}$ are creation and annihilation operators of the radiation field which satisfy the usual commutation relation $[\hat{a},\hat{a}^\dagger]=1$ and $\hat{\sigma}^{(i)}_{\nu}$, where $(\nu=x,y,z,\pm)$, are the usual Pauli, raising and lowering spin operators representing the $i$th qubit.

The atoms are assumed to be initially in a pure state and the field is in a coherent state, therefore the initial wave function of the composite system becomes
\begin{equation}
\vert\psi(0)\rangle = [a\; \vert e_{1},e_{2}\rangle + b \; \vert e_{1},g_{2}\rangle + c \; \vert g_{1},e_{2}\rangle + d \; \vert g_{1},g_{2}\rangle]\otimes \vert\alpha\rangle,
\label{eq:8}
\end{equation}
where $a,b,c$ and $d$, are arbitrary complex quantities that satisfy the condition $\vert a \vert^2 + \vert b \vert^2 +\vert c \vert^2 +  \vert d \vert^2 =1$, and $\vert\alpha\rangle$ is the coherent state defined as
\begin{equation}
\vert\alpha\rangle=\sum_{n} Q_{n} \vert n \rangle;   \qquad Q_{n}=\frac{\alpha^{n}}{\sqrt{ n !}} \exp\left(-\frac{\vert \alpha \vert^2}{2}\right),
\label{eq:10}
\end{equation}
where $\vert \alpha \vert^2=\overline{n}$ is the mean photon number and $\vert n \rangle$ are the photon number states, which satisfy the relations: $\hat{a}^{\dagger} \vert n \rangle =\sqrt{n+1}\vert n+1 \rangle$ and $\hat{a} \vert n+1 \rangle = \sqrt{n+1} \vert n \rangle$. The wave function at any time $t$ latter can be written as
\begin{eqnarray}
\nonumber \vert\psi(t)\rangle &=& \sum_{n} [ A_{n}(t)\vert e_{1},e_{2},n\rangle + B_{n+1}(t)\vert e_{1},g_{2},n+1\rangle + C_{n+1}(t) \vert g_{1},e_{2},n+1\rangle \\
&& + D_{n+2}(t)\vert g_{1},g_{2},n+2\rangle ],
\label{eq:psi_t}
\end{eqnarray}
where $ \vert e_{1},e_{2},n\rangle $ is the state in which both of the two atoms are in the excited state and the field has $n$ photons, while $ \vert e_{1},g_{2},n+1\rangle $ is the state in which the first one is in the excited state and the second is in the ground state and the field has $n+1$ photons, etc. 

Rewriting the Hamiltonian (Eq.~(\ref{eq:H})) as 
\begin{equation}
\hat{H}=\hat{H}_{\circ}+\hat{H}_{int}\;,
\label{eq:14}
\end{equation}
where 
\begin{equation}
\hat{H}_{\circ} = \Omega \; \hat{N} + \frac{\Delta}{2} \sum_{i=1,2} \; \hat{\sigma}^{(i)}_{z}\;,
\label{eq:15}
\end{equation} 
\begin{equation}
\hat{H}_{int}= \lambda_1 \sum_{i=1,2} \; (\hat{a}\hat{\sigma}^{(i)}_{+}+\hat{a}^\dagger \hat{\sigma}^{(i)}_{-})\: + \lambda_2 \; (\hat{\sigma}^{(1)}_{-}\hat{\sigma}^{(2)}_{+}+\hat{\sigma}^{(1)}_{+}\hat{\sigma}^{(2)}_{-})  + \; J \sigma^{(1)}_z \sigma^{(2)}_z ,
\label{eq:16}
\end{equation}
where $\Delta=\omega_{\circ}-\Omega$ is the detuning parameter and $\hat{N}= \hat{a}^\dagger \hat{a} + \frac{1}{2} \sum_{i=1,2} \; \hat{\sigma}^{(i)}_{z}\;,$ is the total number of excitations in the system, which is a constant of motion.
It is more convenient to work in the interaction picture, where we define $\hat{V}_I=\hat{U} \hat{H}_{int} \hat{U}^{\dagger}$ and $\hat{U}=e^{\textrm{i}\hat{H}_{\circ}t}$. This yields
\begin{eqnarray}
\hat{V}_{I}(t) = \; \lambda_1 \sum_{k=1,2}\; (\hat{a}\;e^{\textrm{i}\Delta t} \hat{\sigma}^{(k)}_{+} + \hat{a}^\dagger \; e^{-\textrm{i}\Delta t} \hat{\sigma}^{(k)}_{-})+ \: \lambda_2 ( \hat{\sigma}^{(1)}_{-}\hat{\sigma}^{(2)}_{+}
+ \hat{\sigma}^{(1)}_{+}\hat{\sigma}^{(2)}_{-}) 
 + \; J \sigma^{(1)}_z \sigma^{(2)}_z \;.
\label{eq:V_int}
\end{eqnarray}
Now substituting $\vert\psi(t)\rangle$ and $V_I(t)$ into Schr{\"o}dinger equation
\begin{equation}
\textrm{i} \; \frac{\partial}{\partial t}\vert\psi(t)\rangle = \hat{V}_I(t)\vert\psi(t)\rangle, 
\label{eq:Sch_eqn}
\end{equation}
we get the following system of differential equations
\begin{eqnarray}
\nonumber \textrm{i}\dot{A}_{n}(t) &=& \alpha\; e^{\textrm{i}\Delta t} \;(B_{n+1}(t) + C_{n+1}(t)) + J A_n(t),\\
\nonumber \textrm{i}\dot{B}_{n+1}(t) &=& \alpha \;e^{-\textrm{i} \Delta t }\;A_{n}(t) + \beta \;e^{\textrm{i}\Delta t}\;D_{n+2}(t) + \lambda_{2}\;C_{n+1}(t) - J B_{n+1}(t),\\
\nonumber \textrm{i}\dot{C}_{n+1}(t) &=& \alpha\;e^{-\textrm{i} \Delta t }\;A_{n}(t) + \beta \;e^{\textrm{i}\Delta t}\;D_{n+2}(t) + \lambda_{2}\;B_{n+1}(t) - J C_{n+1}(t),\\
\textrm{i}\dot{D}_{n+2}(t)&=& \beta\; e^{-\textrm{i}\Delta t}\;(B_{n+1}(t) + C_{n+1}(t)) + J D_{n+2}(t),
\label{eq:DE_sys1}
\end{eqnarray}
where $\alpha=\lambda_{1}\sqrt{n+1}$ and $\beta=\lambda_{1}\sqrt{n+2}$. 

Following the same approach that we implemented in our previous work \cite{Sadiek:2019}, and after some calculations, the solution takes the form
\begin{eqnarray}
\nonumber A_{n}(t)&=& A_{n}(0) e^{-\textrm{i}Jt} -\textrm{i}\alpha \sum^{3}_{j=1}[\frac{\delta_{j}}{m_{j}+\textrm{i}(\Delta+J})( e^{(m_{j}+\textrm{i}\Delta )t}- e^{-\textrm{i}Jt})],\\
\nonumber B_{n+1}(t)&=&\frac{1}{2}[(B_{n+1}(0)-C_{n+1}(0))e^{\textrm{i}(\lambda_{2}+J)t}+\sum^{3}_{j=1}\delta_{j} e^{m_{j}t}],\\
\nonumber C_{n+1}(t)&=&\frac{1}{2}[(C_{n+1}(0)-B_{n+1}(0))e^{\textrm{i}(\lambda_{2}+J)t}+\sum^{3}_{j=1}\delta_{j} e^{m_{j}t}],\\
D_{n+2}(t)&=& D_{n+2}(0) e^{-\textrm{i}Jt} -\textrm{i}\beta \sum^{3}_{j=1}[\frac{\delta_{j}}{m_{j}-\textrm{i}(J-\Delta)}( e^{(m_{j}-\textrm{i}\Delta )t}-e^{-\textrm{i}Jt})],
\label{eq:coef_soln}
\end{eqnarray}
where
\begin{eqnarray}
\nonumber \delta_{1}&=&(B_{n+1}(0)+C_{n+1}(0))-(\delta_{2}+\delta_{3}),\\
\nonumber \delta_{2}&=&\frac{1}{(m_{1}-m_{2})(m_{3}-m_{2})} \lbrace 2\alpha A_{n}(0)[\textrm{i}(m_{1}+m_{3})-\lambda_{2} -\Delta]+2\beta D_{n+2}(0) [\textrm{i}(m_{1}+m_{3})\\
\nonumber &&- \lambda_{2}+\Delta]+ [\textrm{i}(m_{1}+m_{3})(\lambda_{2} -J
-\textrm{i}m_{1})-2(\alpha^{2}+\beta^{2})-(\lambda_{2}-J)^2-m_{1}^{2}]\\
\nonumber && \times(B_{n+1}(0)+C_{n+1}(0))\rbrace, \\
\nonumber \delta_{3}&=&\frac{1}{(m_{1}-m_{3})(m_{2}- m_{3})}\lbrace 2\alpha A_{n}(0)[\textrm{i}(m_{1}+m_{2})-\lambda_{2}-\Delta]+2\beta D_{n+2}(0)[\textrm{i}(m_{1}+m_{2})\\
\nonumber &&-\lambda_{2}+\Delta]+[\textrm{i}(m_{1}+m_{2})(\lambda_{2}-J-\textrm{i}m_{1})-2(\alpha^{2}+\beta^{2})-(\lambda_{2}-J)^{2}-m_{1}^{2}]\\
&& \times(B_{n+1}(0)+C_{n+1}(0))\rbrace,
\label{eq:24}
\end{eqnarray}
and 
\begin{eqnarray}
\nonumber m_{1}&=&(v_{1}+v_{2})-\textrm{i}\frac{\lambda_{2}+J}{3},\\
\nonumber m_{2}&=&-\frac{v_{1}+v_{2}}{2}+\textrm{i} \frac{\sqrt{3}}{2}(v_{1}-v_{2})-\textrm{i}\frac{\lambda_{2}+J}{3},\\
m_{3}&=&-\frac{v_{1}+v_{2}}{2}-\textrm{i} \frac{\sqrt{3}}{2}(v_{1}-v_{2})-\textrm{i}\frac{\lambda_{2}+J}{3},
\label{eq:ms}
\end{eqnarray}
where
\begin{equation}
v_{1}=[ -\frac{\mu}{2}+(\frac{\mu^{2}}{4}+\frac{\eta^{3}}{27})^\frac{1}{2}]^\frac{1}{3}; \qquad
v_{2}=[-\frac{\mu}{2}-(\frac{\mu^{2}}{4}+\frac{\eta^{3}}{27})^\frac{1}{2}]^\frac{1}{3},
\label{eq:v1_v2}
\end{equation}
and 
\begin{eqnarray}
\nonumber \mu &=& \frac{2}{27} \textrm{i} [\left(-27 \alpha ^2 \Delta +27 \beta ^2 \Delta +8 J^3+18 \alpha ^2 J+18 \beta ^2 J-18 \Delta ^2 J\right) \\
\nonumber &+&  \lambda_2  \left(-9 \alpha ^2-9 \beta ^2+9 \Delta ^2-12 J^2\right)+6 J \lambda_2 ^2-\lambda_2 ^3],
\label{eq:mu}
\end{eqnarray}
\begin{equation}
 \eta=2 \left(\alpha ^2+\beta ^2\right)+\Delta ^2+\frac{1}{3} (\lambda_2 -2 J)^2\;.
\label{eq:eta}
\end{equation}
The initial values of the coefficients are as follows 
\begin{eqnarray}
A_{n}(0)= Q_{n} \;a,\;\; B_{n+1}(0)=Q_{n+1} \;b,\;\; C_{n+1}(0)=Q_{n+1} \;c,
\; \; D_{n+2}(0)= Q_{n+2} \;d.
\label{eq:init_values}
\end{eqnarray}
Now we can construct the system wave function $\vert\psi(t)\rangle$ and calculate the system density matrix $\hat{\rho}(t)=\vert\psi(t)\rangle\langle\psi(t)\vert$. The reduced density matrix of the two atoms, $\hat{\rho}_{\textrm{red}}(t)$, can be obtained by tracing out the field 
\begin{equation}
\hat{\rho}_{\textrm{red}}(t)=\textrm{Tr}_{\textrm{field}}\; \hat{\rho}(t)= \sum_{l} \langle l \vert \psi(t)\rangle \langle \psi(t) \vert l \rangle.
\label{eq:qs_rdm}
\end{equation}

\section{Time evolution of entanglement and population inversion}
Utilizing the derived analytic solution, we can investigate the entanglement dynamics between the two atoms and their population inversion starting from different initial states that lead to ESD.
For convenience, we set $\hbar=1$, $\lambda_1=1$ and represent the other parameters ($\lambda_2$, $J$ and $\Delta$) in units of $\lambda_1$. The entanglement between the two quantum systems is quantified using the concurrence function $C(\rho_{\textrm{red}})$ \cite{Wootters:1998}. The concurrence is related to the entanglement of formation $E_{f}$ through the formula $E_{f}(\rho_{\textrm{red}})= \mathcal{E}(C(\rho_{\textrm{red}}))$,
where $\mathcal{E}$ is defined as
\begin{equation}
\mathcal{E}(C(\rho_{\textrm{red}}))= h\left(\frac{1+ \sqrt{1-C^{2}(\rho_{\textrm{red}})}}{2} \right),
\label{eq:49}
\end{equation}
$h$ is the Shannon entropy function
$h(x)=-x \log_{2} x - (1-x)\log_{2} (1-x)$, and the concurrence can be calculated from
$C(\rho_{\textrm{red}})= \max\;[0,\varepsilon_{1}-\varepsilon_{2}-\varepsilon_{3}-\varepsilon_{4}]$.
The $\varepsilon_{i}$ arranged in decreasing order are the square root of the four eigenvalues of the non-Hermitian matrix $R\equiv \rho_{\textrm{red}}\tilde{\rho}_{\textrm{red}}$,
Where $\tilde{\rho}_{\textrm{red}}=(\hat{\sigma}_{y}\otimes\hat{\sigma}_{y})\rho^{*}_{\textrm{red}}(\hat{\sigma}_{y}\otimes\hat{\sigma}_{y})$. 
Both of  $C(\rho_{\textrm{red}})$ and $E_{f}(\rho_{\textrm{red}})$ take values from $0$ for a separable (disentangled) state to $1$ for a maximally entangled state. On the other hand, atomic population inversion is defined as the difference between the probabilities of finding the atom in its excited state and ground state or simply the expectation value of the operator $\hat{\sigma}_{z}$. Using the reduced density matrix of any one of the two identical atoms $\hat{\rho}_{1}(t)$, which can be obtained by tracing out the other one in $\hat{\rho}_{red}$ (Eq.~\ref{eq:qs_rdm}), we can evaluate $\langle \hat{\sigma}_{z}(t) \rangle$ as  
\begin{equation}
\langle \hat{\sigma}_{z}(t) \rangle = Tr [\hat{\rho}_{1}(t)\hat{\sigma}_{z}]
= \sum_{n=0}^{\infty} \vert A_{n}(t) \vert^2 +  \vert B_{n+1}(t)\vert^2 - \vert C_{n+1}(t) \vert^2 - \vert D_{n+2}(t) \vert^2.
\label{eq:sigma_z}
\end{equation}
\subsection{Initial Bell state}
\begin{figure}[htbp]
 \centering
\subfigure{\includegraphics[width=6.4cm]{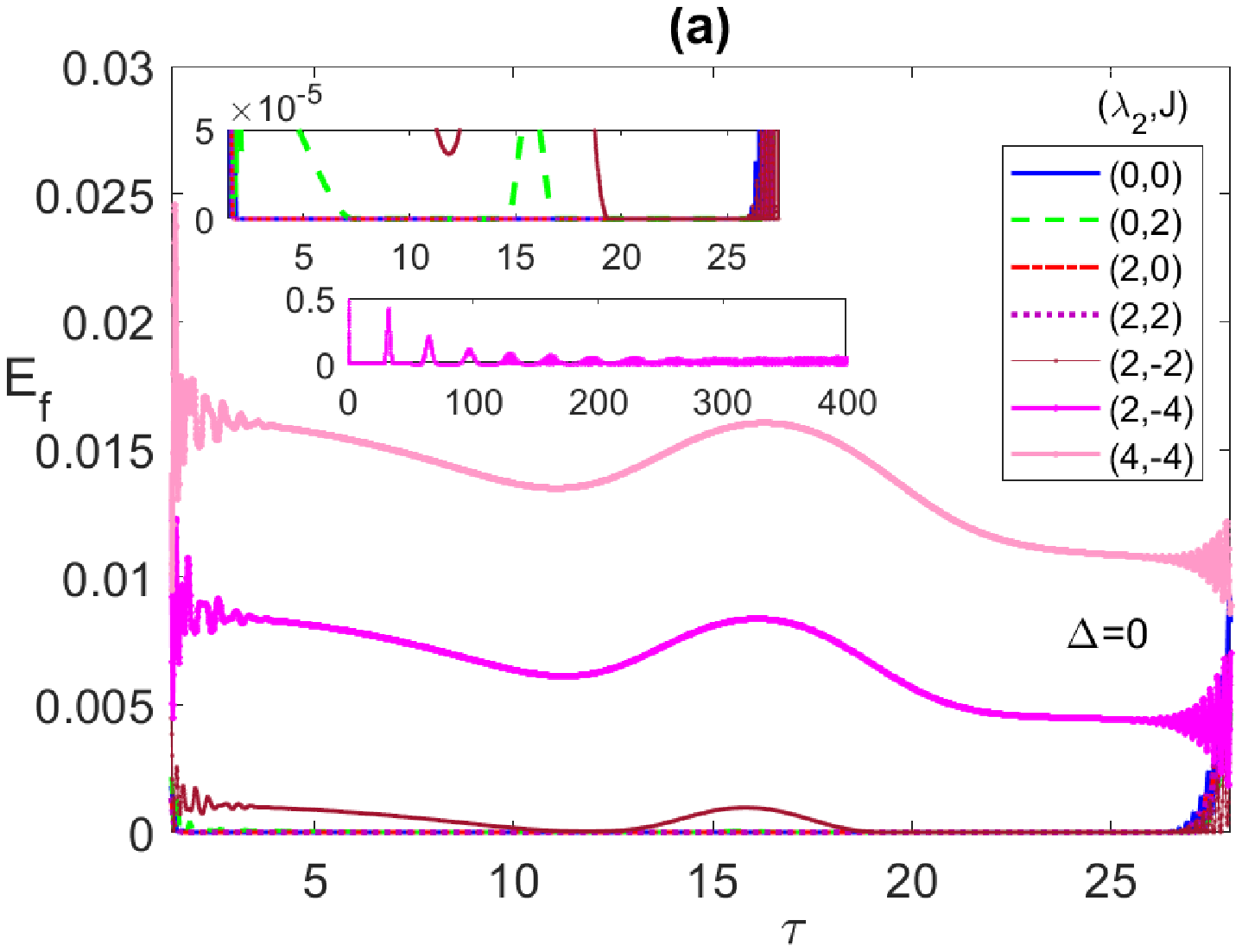}}\quad 
\subfigure{\includegraphics[width=6.4cm]{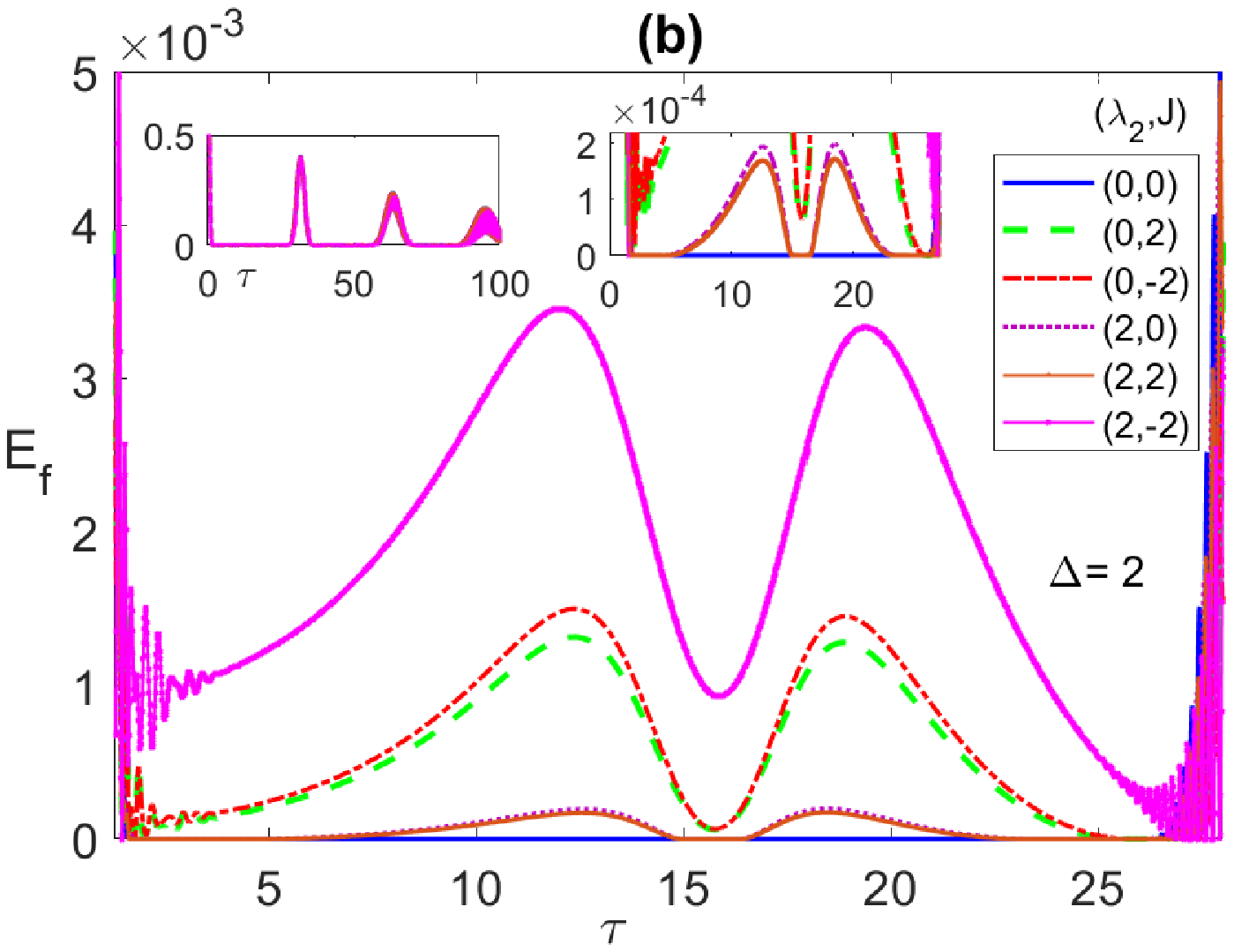}}\\
\subfigure{\includegraphics[width=6.4cm]{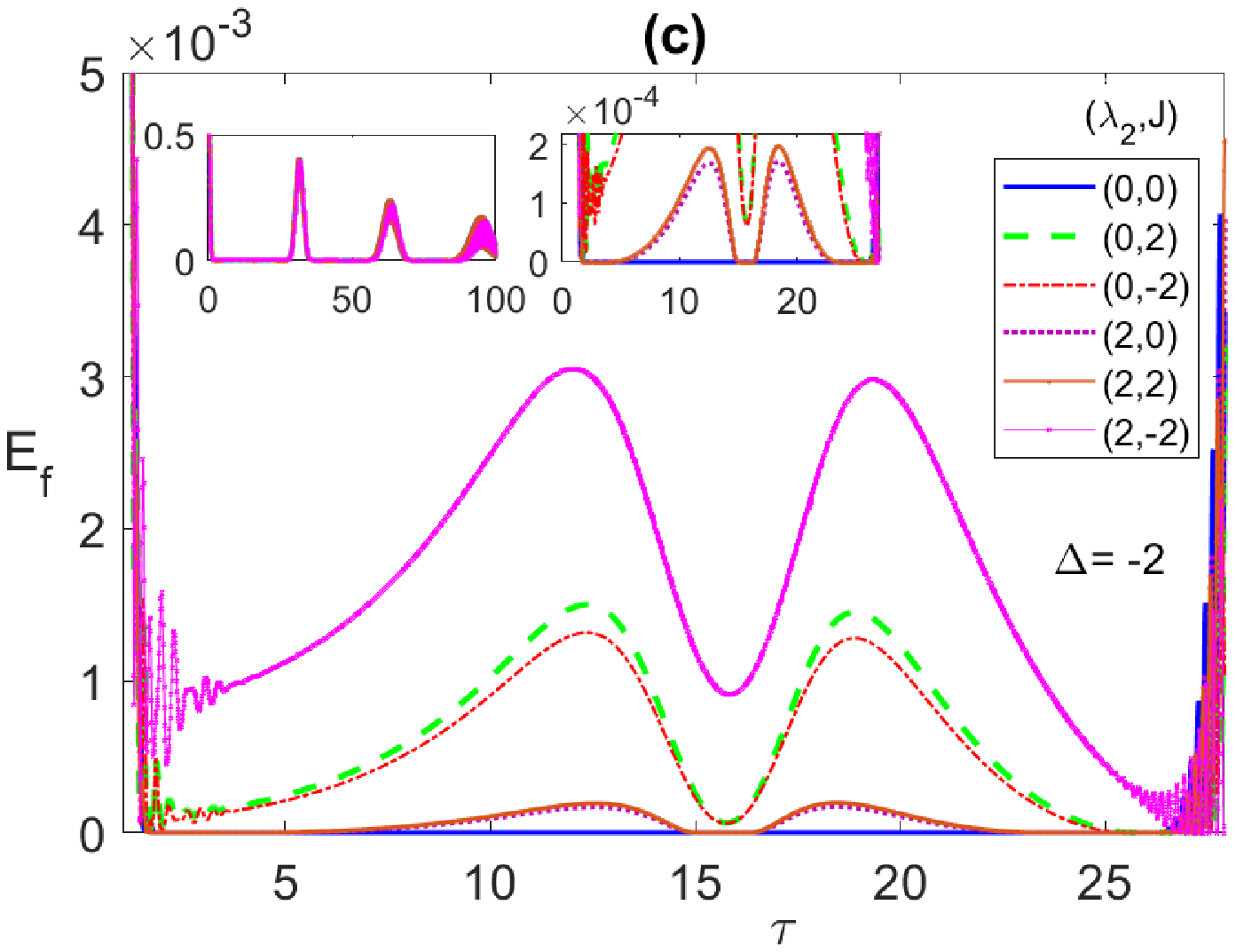}}\quad
\subfigure{\includegraphics[width=6.4cm]{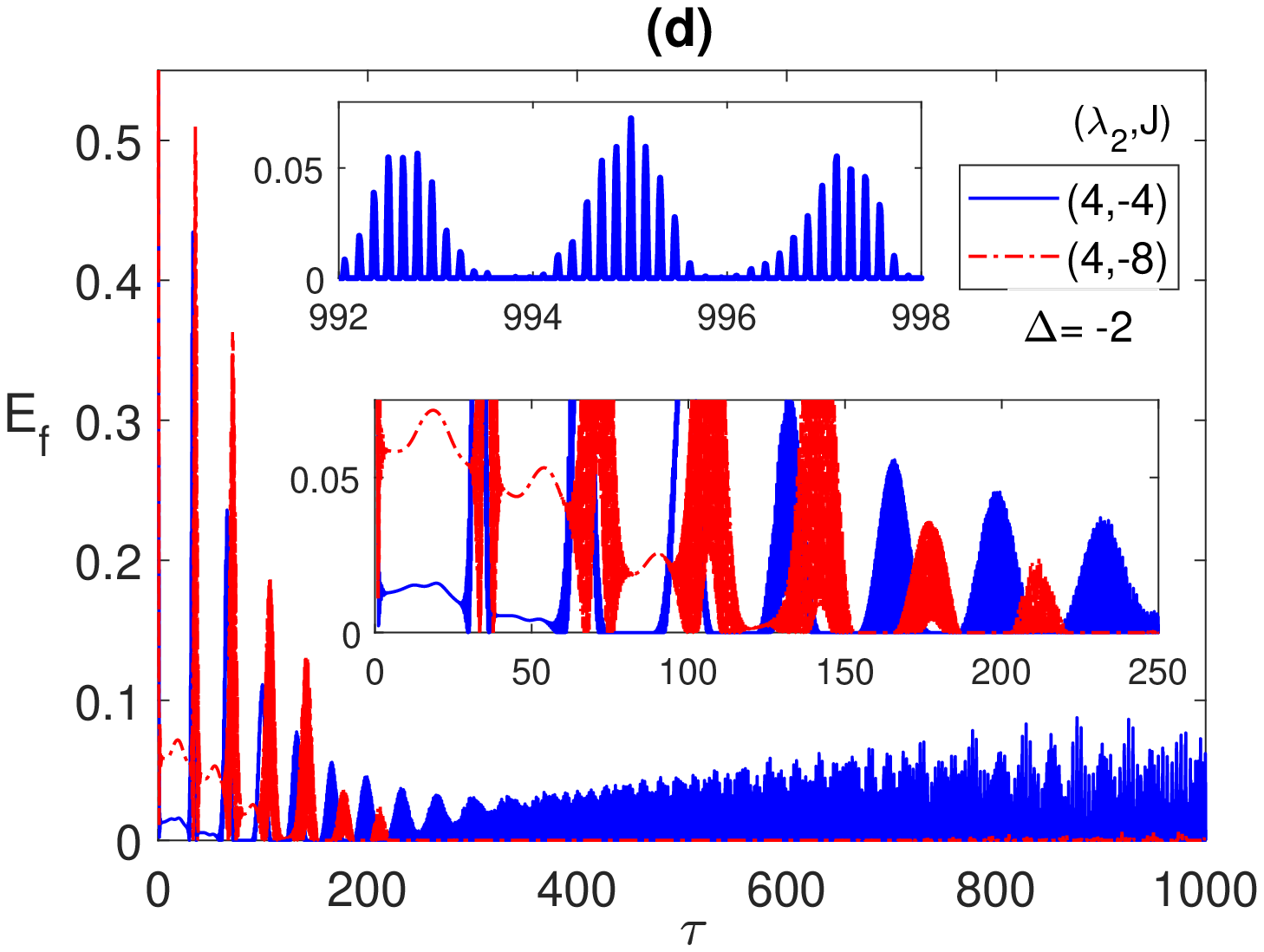}}\\
\caption{{
The entanglement $E_f$ versus the scaled time $\tau=\lambda_1 t$ with the two atoms are initially in a correlated Bell state $\psi_{Bc}=(\vert e_{1}\rangle \vert e_{2}\rangle+\vert g_{1}\rangle \vert g_{2}\rangle)/\sqrt{2}$ and the field is in a coherent state with mean number of photons $\bar{n}=100$, where the detuning parameter $\Delta$ takes the values (a) 0; (b) $2$; (c) $-2$; (d) $-2$, while the dipole-dipole and Ising coupling parameters $(\lambda_2, J)$ take various values as shown in the legend in each panel.}}
\label{fig1}
\end{figure}

In Fig.~\ref{fig1}, we explore the dynamics of entanglement and population inversion, in terms of the scaled time $\tau=\lambda_1 t$, starting from an initial correlated Bell state $\psi_{Bc}=(\vert e_{1}\rangle \vert e_{2}\rangle+\vert g_{1}\rangle \vert g_{2}\rangle)/\sqrt{2}\;$ with the radiation field is in a coherent state with field intensity corresponding to $\bar{n}=100$. In the forthcoming discussion, we set $\bar{n}=100$ everywhere except when otherwise is mentioned explicitly. Starting from such an initial state the system shows ESD, where the entanglement changes abruptly from a non-zero to an exact zero value, as illustrated in the different panels of the figure.

In our forthcoming discussion and in the legends of all plots, we will use the pair notation $(a,b)$ to refer to the values of the two parameters $\lambda_2$ and $J$, where $"a"$ refers to the value of $\lambda_2$, while $"b"$ refers to the value of $J$.
In Fig.~\ref{fig1}(a), we study the entanglement dynamics at resonance, i.e. zero detuning $(\Delta=0)$, at different choices of $\lambda_2$ and $J$. As can be noticed, when the two parameters are $(0,0)$, i.e. non-interacting atoms, there is an ESD for a whole period of time from about 2 to 27 before reviving again, represented by the solid (blue) line. 
Turning on the Anti-Ferromagnetic (AF) Ising coupling, $J>0$, with a small value such as $1$ shows a negligible effect on the ESD, but increasing that value to $2$, shows a partial removal of the ESD, as shown clearly in the upper inner inset, dashed (green) line. When the Ferromagnetic (FM) Ising coupling, $J<0$, is considered instead, it shows the same exact behavior as the Ferromagnetic case for this initial state at $\lambda_2$=0. Turning on the dipole coupling $\lambda_2$ at $J=0$, i.e for (2,0), shows no effect on ESD at small values, such as 2 represented by the dot-dashed (red line), however applying higher values of $\lambda_2$, removes the ESD partially and then eliminates it totally as it increases. Interestingly, while we could remove the ESD completely by setting the values at (4,0), turning on AM Ising even at a small value is devastating for the entanglement, (4,1) for instance, shows a complete ESD. However, applying an FM Ising proves to be an enhancement for the entanglement, where (2,-2) removes ESD partially, represented by the (brown) solid line with dot marks, while (2,-4) (the (magenta) solid line with x marks) and (4,-4) (the (pink) solid line with star marks) remove the ESD completely and boost the entanglement value considerably. This means, at resonance, combining $\lambda_2$ and FM $J$ is very useful in removing ESD and more effective than any of them separately, while combining $\lambda_2$ and AF $J$ cancels out their individual effects and maintains the ESD.

\begin{figure}[htbp]
 \centering
\subfigure{\includegraphics[width=10cm]{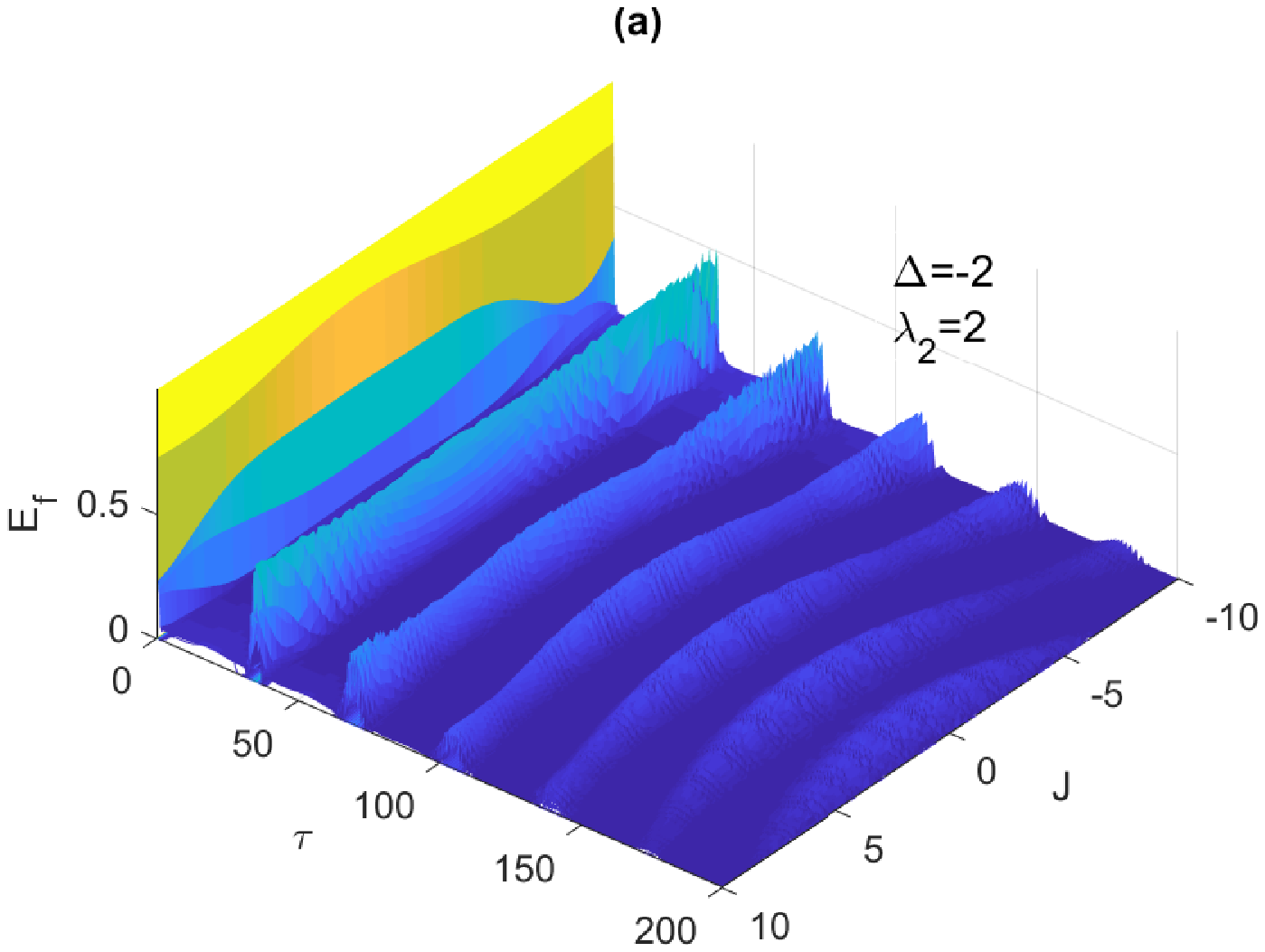}}\\ 
\subfigure{\includegraphics[width=9cm]{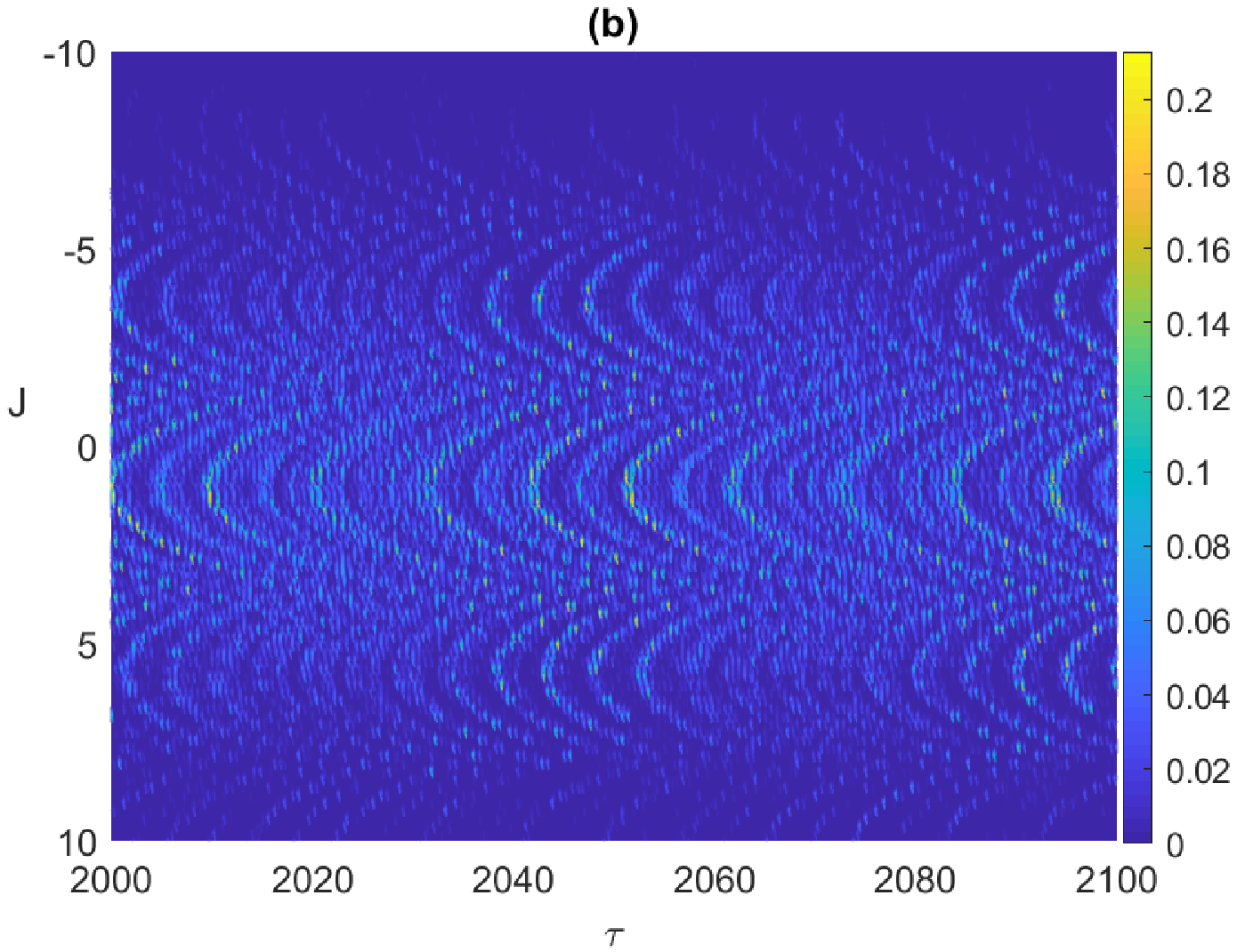}}\\
\subfigure{\includegraphics[width=9cm]{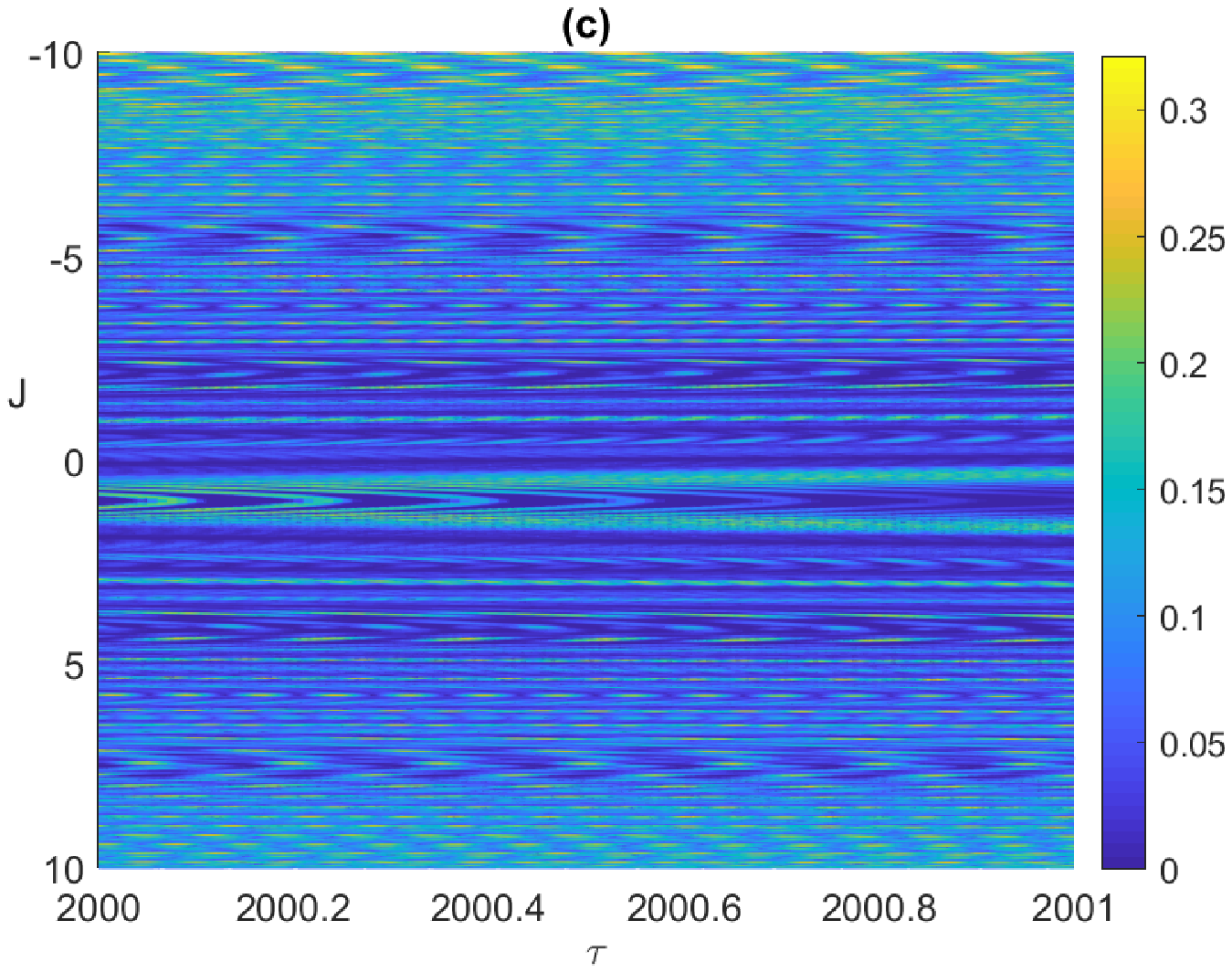}}\\
\caption{
The entanglement $E_f$ versus the scaled time $\tau=\lambda_1 t$ and the Ising coupling J
with the two atoms are initially in the correlated Bell state $\psi_{Bc}$ and the field is in a coherent state, where $\lambda_2=2$ and $\Delta=-2$, at early time in (a) and asymptotically (contour plot) in (b), and asymptotically (contour plot) for the initial anti-correlated Bell state $\psi_{Ba}=(\vert g_{1}\rangle \vert e_{2}\rangle+\vert e_{1}\rangle \vert g_{2}\rangle)/\sqrt{2}$ in (c).}
\label{fig2}
\end{figure}

In Fig.~\ref{fig1}(b), we consider an out of resonance case, with positive detuning $(\Delta=2)$. For non-interacting atoms, with parameter values (0,0), the ESD sustains over the entire time period (blue line) from about 2 to 27. However, turning on FM (or AF) Ising has a strong impact on the ESD compared with the $\Delta=0$ case, where setting the parameters at $(0,\pm 2)$ removes completely the ESD with the AM entanglement peaks (red line) is slightly higher than that of the FM one (green line). Turning on the dipole interaction while keeping $J=0$, removes the ESD partially, for (2,0), and for higher values of $\lambda_2$, it completely eliminates the ESD and enhances the Entanglement. Adding the AF Ising to the dipole interaction, (2,2), does not have a large impact on the entanglement behavior except for a slight shift (brown line). However, adding the FM Ising removes the ESD and enhances the entanglement significantly, even for parameter values as low as (2,-2) (magenta line). Increasing the value of (FM) $J$ further increases the entanglement considerably (not shown here). The entanglement dynamics for negative detuning, $\Delta=-2$, illustrated in Fig.~\ref{fig1}(c), does not exhibit notable differences from the positive detuning case, $\Delta=2$, except the peaks (green line) for (0,2) is slightly higher, this time, than that of (0,-2) (red line), and so is (2,2) compared to (2,0). It is essential to investigate the effect of the different system parameters on the asymptotic behavior of the entanglement between the two atoms, which is considered in Fig.~\ref{fig1}(d). While, as we concluded from Fig.~\ref{fig1}(a), (b) and (c), combining the dipole-dipole and FM Ising interactions is very effective in eliminating the ESD and enhancing the entanglement in the first period, it is clear from panel (d) that setting the parameter values even at (4,-4) can remove the ESD from the first and second periods but not the latter ones. In fact, the entanglement exhibits asymptotically an irregular oscillation interrupted with a persistent ESD small periods, as shown in the upper inset of panel (d). Applying higher values of the parameters, such as (4,-8) (green line), may remove the ESD in more early periods, as shown in the lower inset of panel(d), however, it causes the entanglement to vanish entirely asymptotically.

\begin{figure}[htbp]
 \centering
  \subfigure{\includegraphics[width=6.5cm]{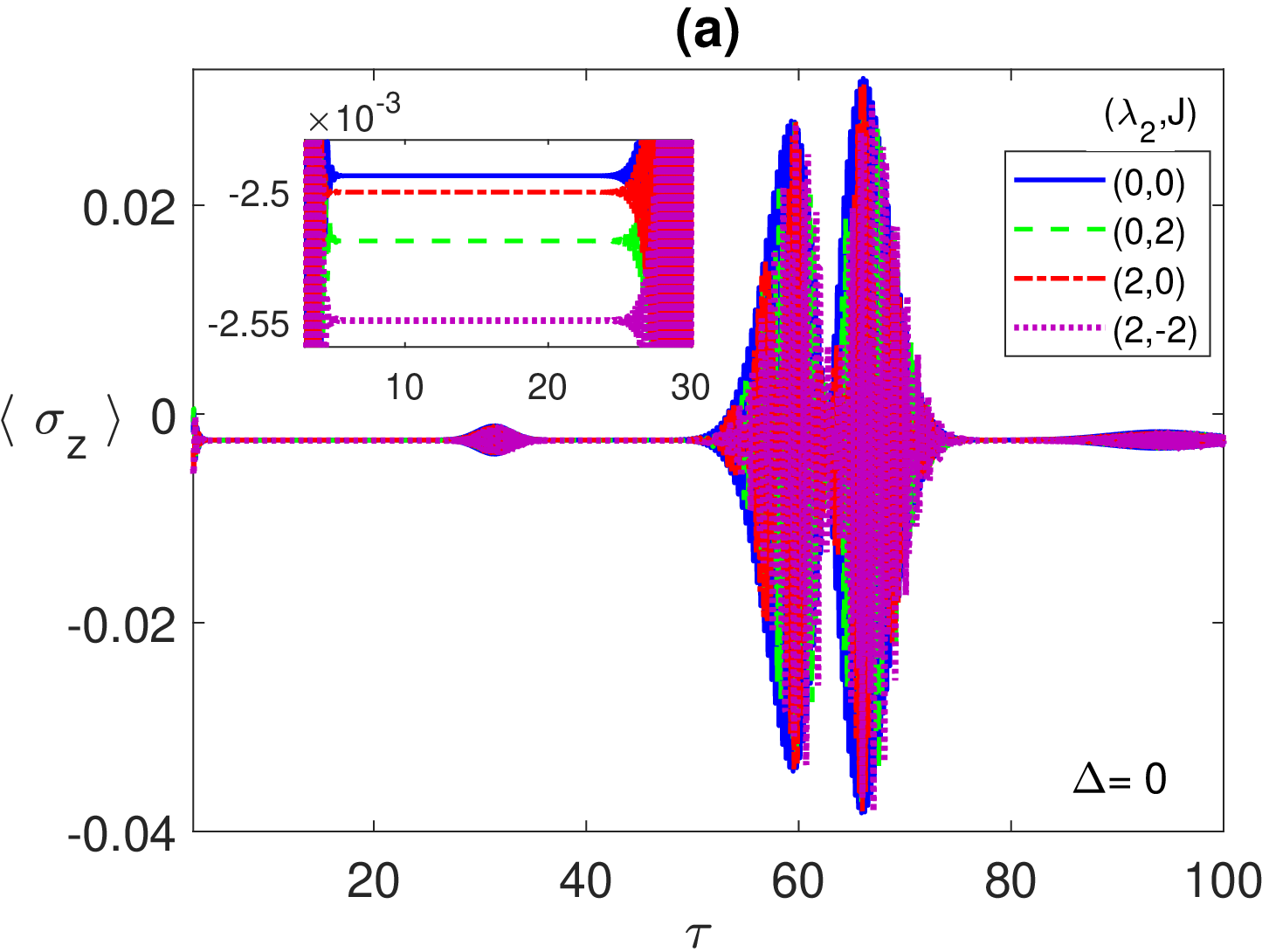}}\quad 
  \subfigure{\includegraphics[width=6.4cm]{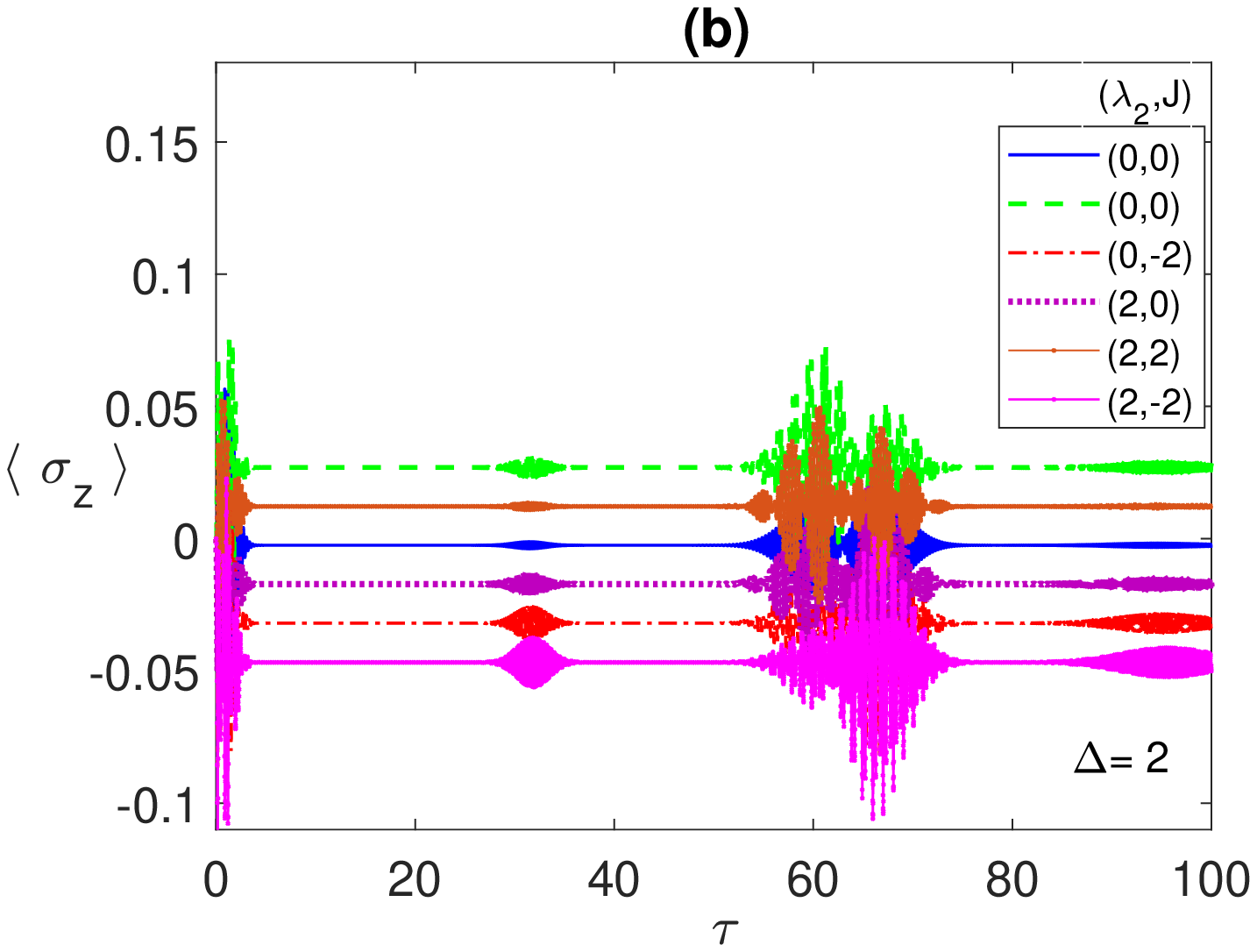}}\\
  \subfigure{\includegraphics[width=6.5cm]{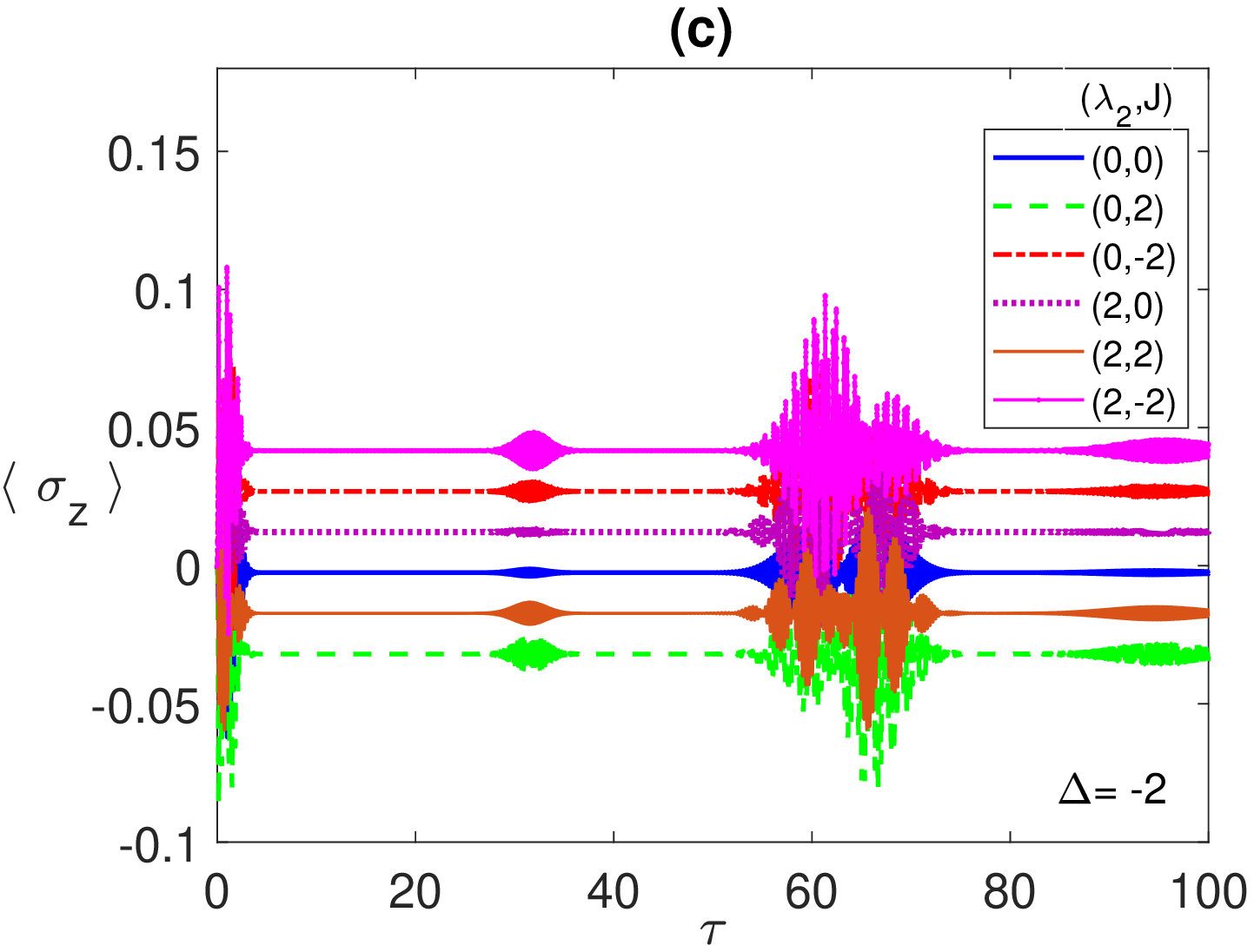}}\quad
  \subfigure{\includegraphics[width=6.4cm]{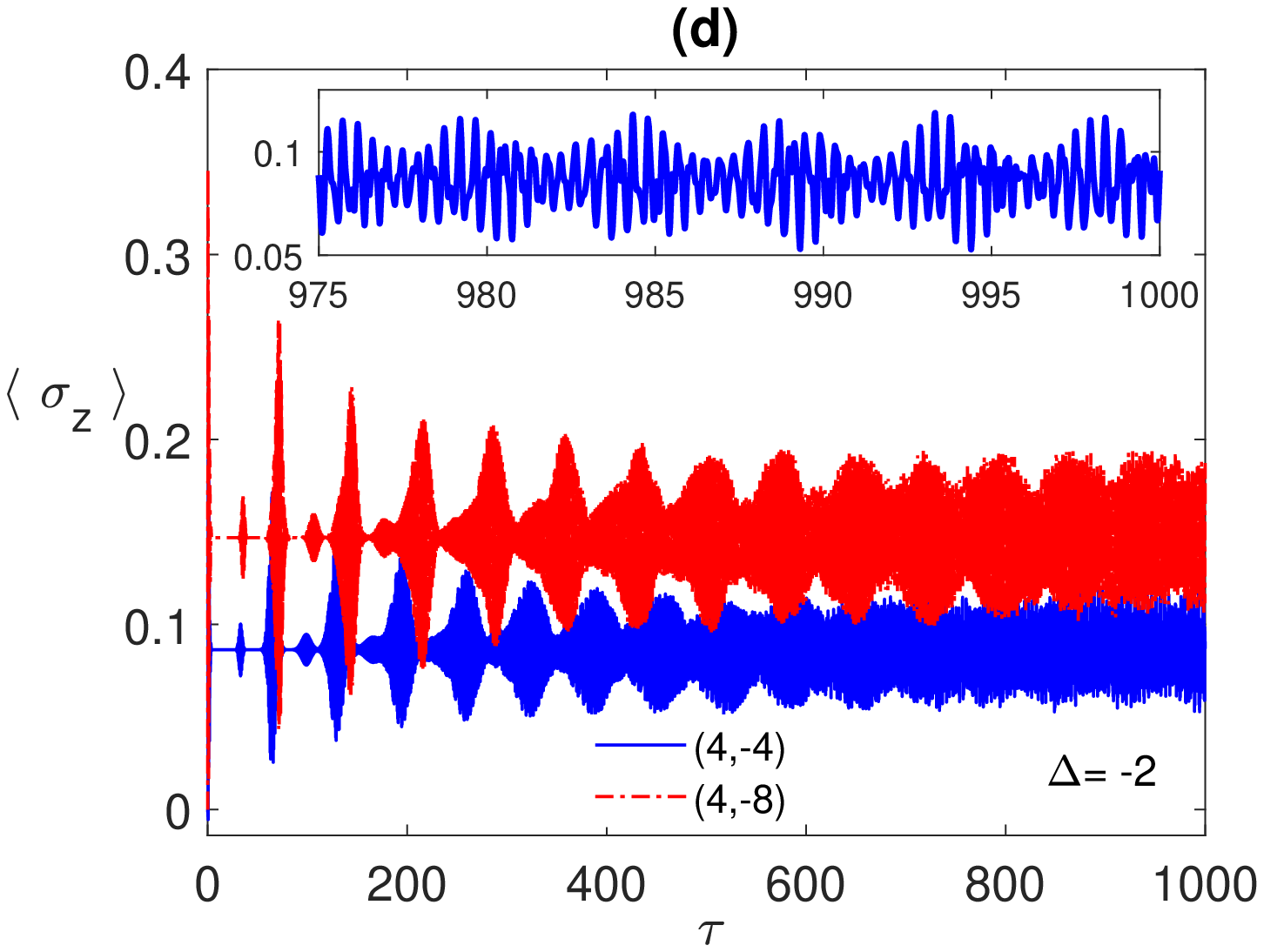}}\\
\caption{{
The population inversion $\langle \sigma_z \rangle$ versus the scaled time $\tau=\lambda_1 t$ with the two atoms are initially in a correlated Bell state $\psi_{Bc}=(\vert e_{1}\rangle \vert e_{2}\rangle+\vert g_{1}\rangle \vert g_{2}\rangle)/\sqrt{2}$ and the field is in a coherent state with $\bar{n}=100$, where in (a) $\Delta=0$; (b) $2$; (c) $-2$; (d) $-2$, while $(\lambda_2, J)$ take various values as shown in the legend in each panel.}}
\label{fig3}
\end{figure}
It is always very insightful to monitor the system dynamics using 3-dimensional plots, which shows the behavior of the system over a wide continuous range of the parameter values. In Fig.~\ref{fig2}(a) and (b), we depict the time evolution of the entanglement versus the Ising coupling parameter, where $-10\leq J \leq 10$ at early times and asymptotically respectively. As can be noticed, increasing the Ising coupling value may remove the ESD at early times as shown in Fig.~\ref{fig2}(a), but eventually causes the entanglement to vanish completely without revival as illustrated in Fig.~\ref{fig2}(b), where the ESD period length increases with $J$ until eliminating the entanglement completely.

Testing the system dynamics starting from another maximally entangled state, the Bell anti-correlated $\psi_{Ba}=(\vert g_{1}\rangle \vert e_{2}\rangle+\vert e_{1}\rangle \vert g_{2}\rangle)/\sqrt{2}$, showed slight differences but not that significant, in the behavior of the entanglement, ESD, and the atomic population compared with the correlated Bell state at the early times, however, asymptotically the ESD can be removed entirely from the system by increasing the Ising coupling value as illustrated Fig.~\ref{fig2}(c). Interestingly, as can be observed in Fig.~\ref{fig2}(b) and (c), the entanglement behavior is symmetric as the values of $J$ changes from the positive to the negative values, however the symmetry center is not $J=0$ but around $0.9$.

In Fig.~\ref{fig3}, we illustrate the atomic population inversion for one of the two atoms, starting from an initial correlated Bell state $\psi_{Bc}$ with the radiation field is in the coherent state. Under all different conditions, the atomic population shows the usual collapse-revival pattern that is a sign of atom-radiation field interaction. It collapses to a constant value that depends on the chosen set of parameters. 
In Fig.~\ref{fig3}(a), we discuss the resonance case, where, as can be noticed, at the parameter values (0,0), the collapse constant value (blue line) is around $-2.45\times 10^{-3}$ but when $J$ is turned on, $(0,\pm2)$, the constant value shifts down (green line) and the collapse period slightly increases. On the other hand, for the values (2,0) (which coincides with (2,2)), the constant value (red line) gets closer to the (0,0) case. Combining the dipole and FM Ising couplings, (2,-2) (purple line), shows a big shift away from the (0,0) case, which agrees with our observations from the entanglement dynamics. By comparing Fig.~\ref{fig1} and Fig.~\ref{fig3}, it is clear that the collapse periods match the ESD periods, while the revivals synchronize with the entanglement revivals from death, i.e. the entanglement revives from death when the atoms exchange energy with the field.

In Fig.~\ref{fig3}(b) and (c), we discuss the non-zero detuning cases, $\Delta =2$, and $-2$ respectively. Remarkably, there is an inverted symmetry existing when one compares the atomic population in the two cases, where each collapse line at specific parameter values , say (2,-2), is far from the (0,0) line equally for $\Delta=2$ and $-2$ but at the opposite sides. Also, as we pointed in Fig.~\ref{fig3}(a), the relative separation among the different collapse lines matches the corresponding entanglement values relative separation. Comparing the three different detuning cases, in panel (a), (b), and (c), it is clear that the off-resonance condition splits the collapse lines away from each other considerably compared with the resonance case. The asymptotic behavior of the atomic population is illustrated in Fig.~\ref{fig3}(d) at the parameter values (4,-4) (blue line) and (4,-8) (red line). As can be seen, the collapse periods get shorter with time before disappearing asymptotically and the general profile turns into an irregular continuous oscillation with varying reduced amplitude (as shown in the inset), which indicates a continuous energy exchange between the atoms and the field.
\subsection{Partially entangled initial (W) state}
\begin{figure}[htbp]
 \centering
\subfigure{\includegraphics[width=6.4cm]{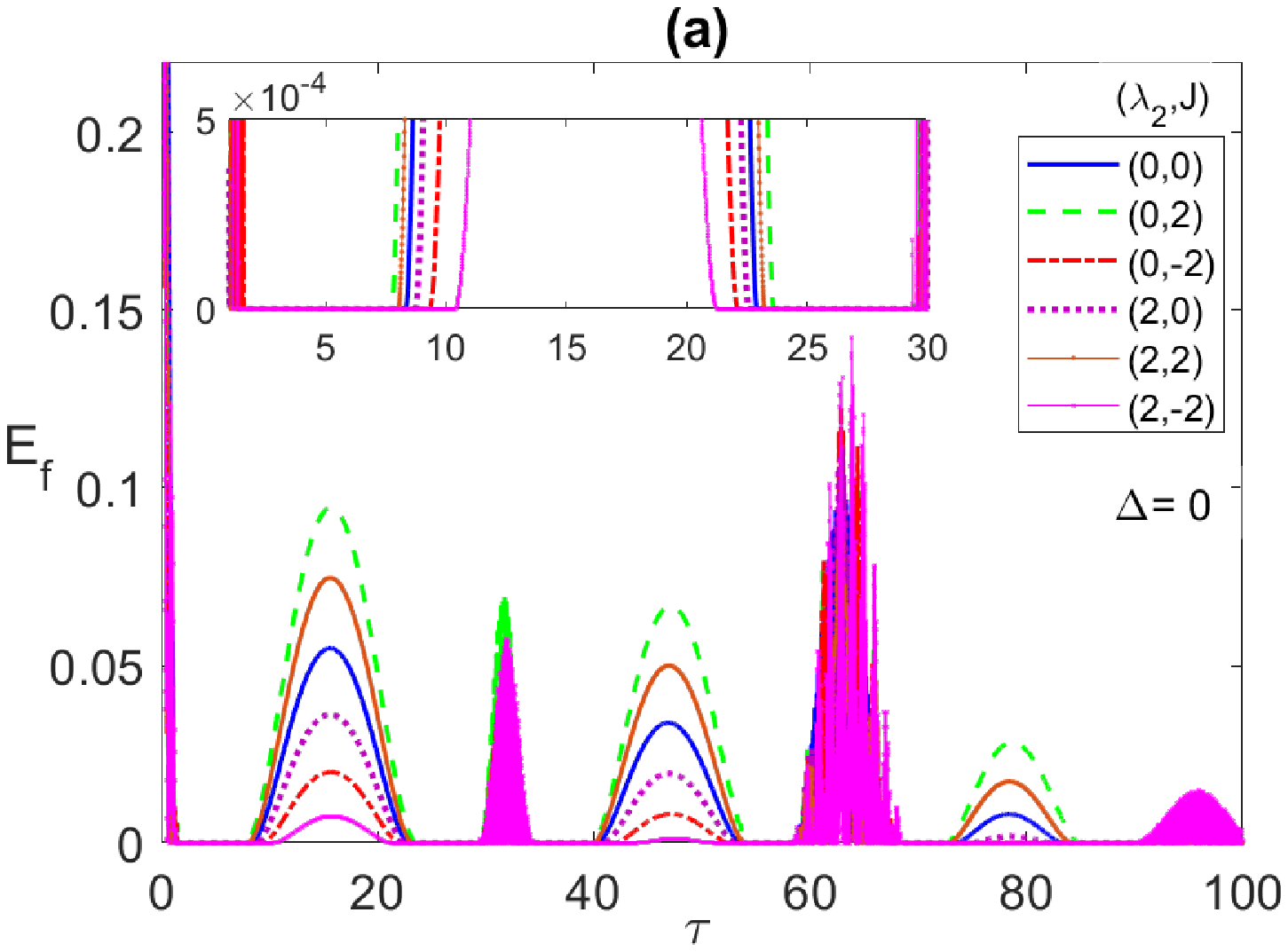}}\quad 
\subfigure{\includegraphics[width=6.4cm]{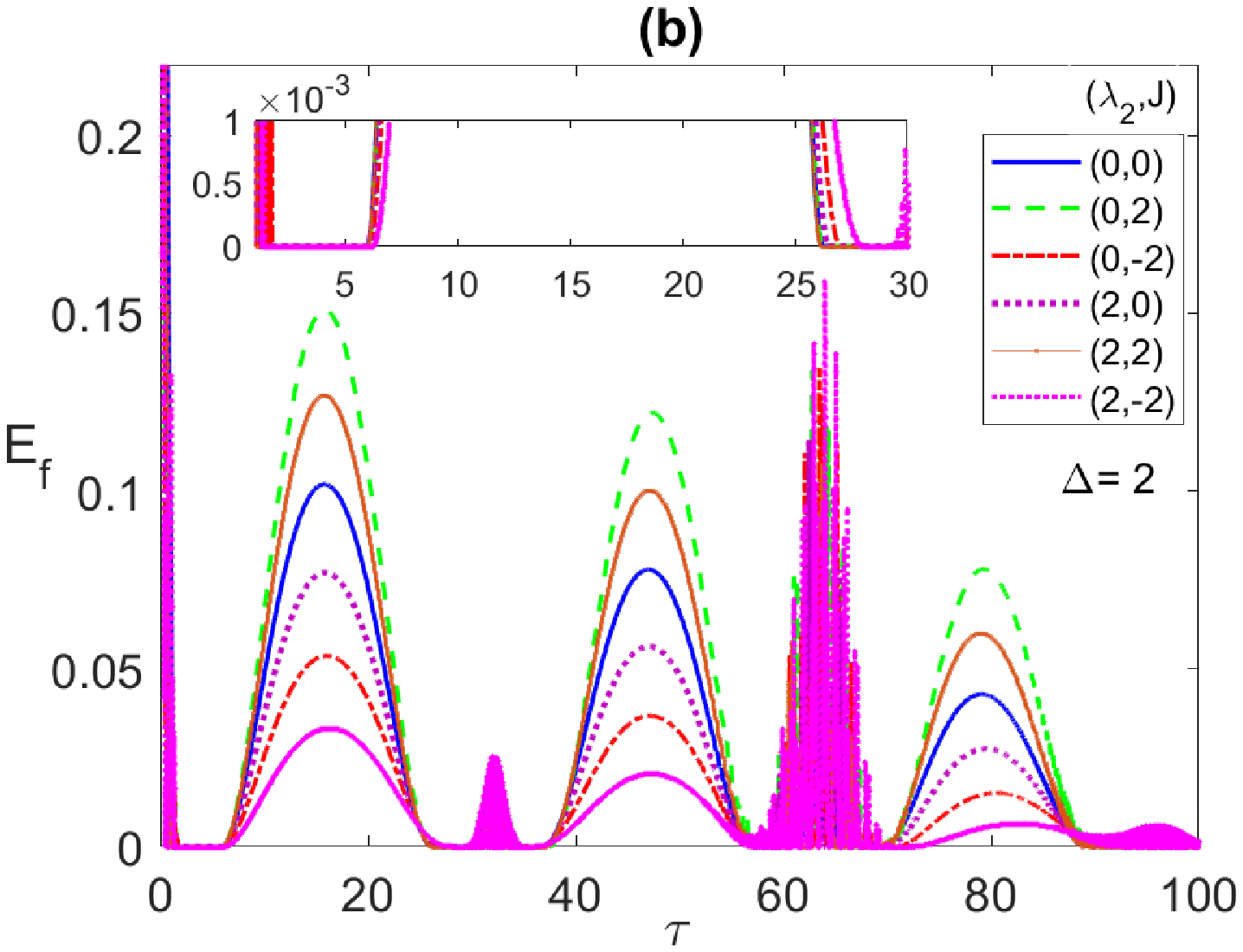}}\\
\subfigure{\includegraphics[width=6.4cm]{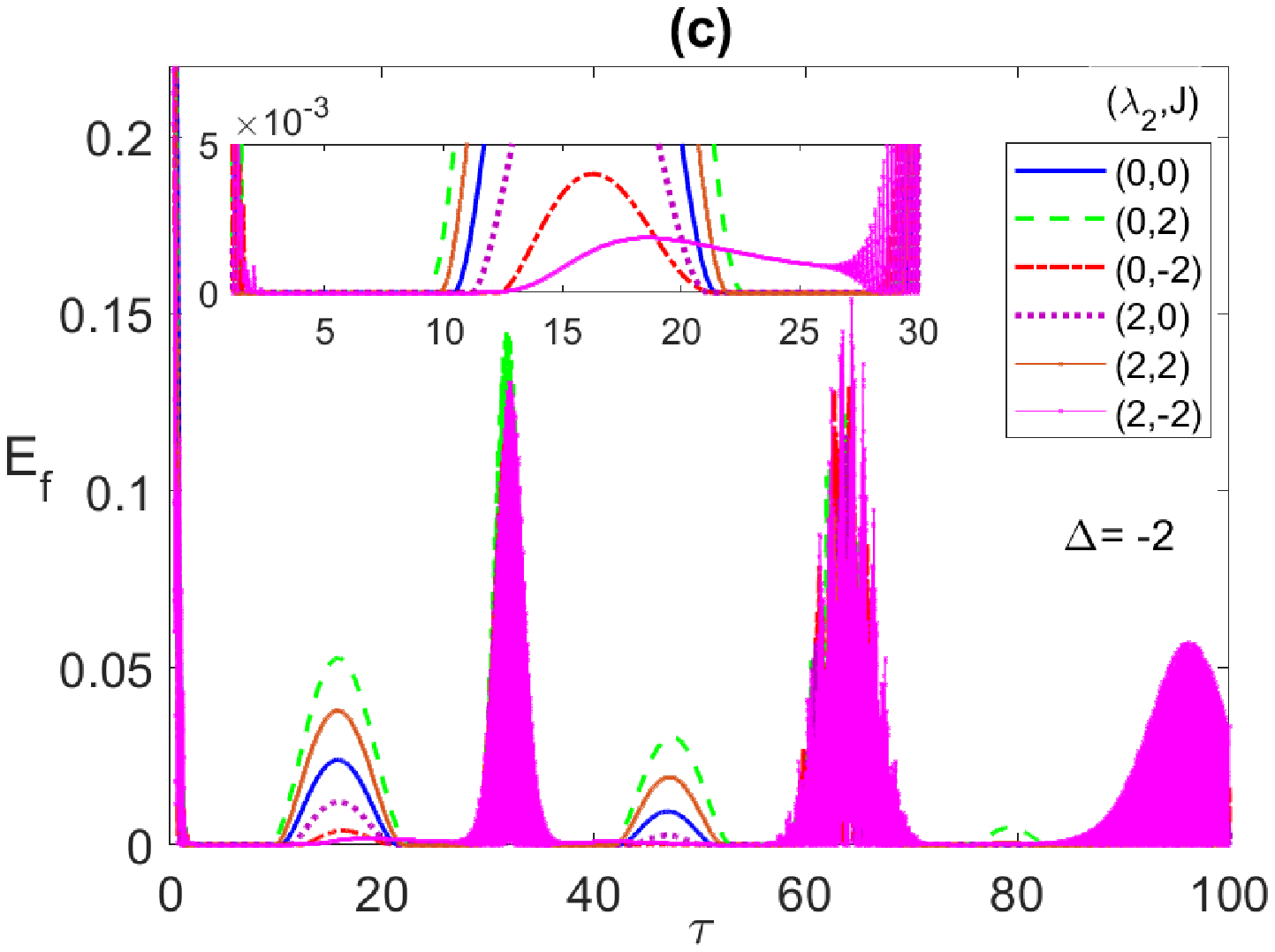}}\quad
\subfigure{\includegraphics[width=6.4cm]{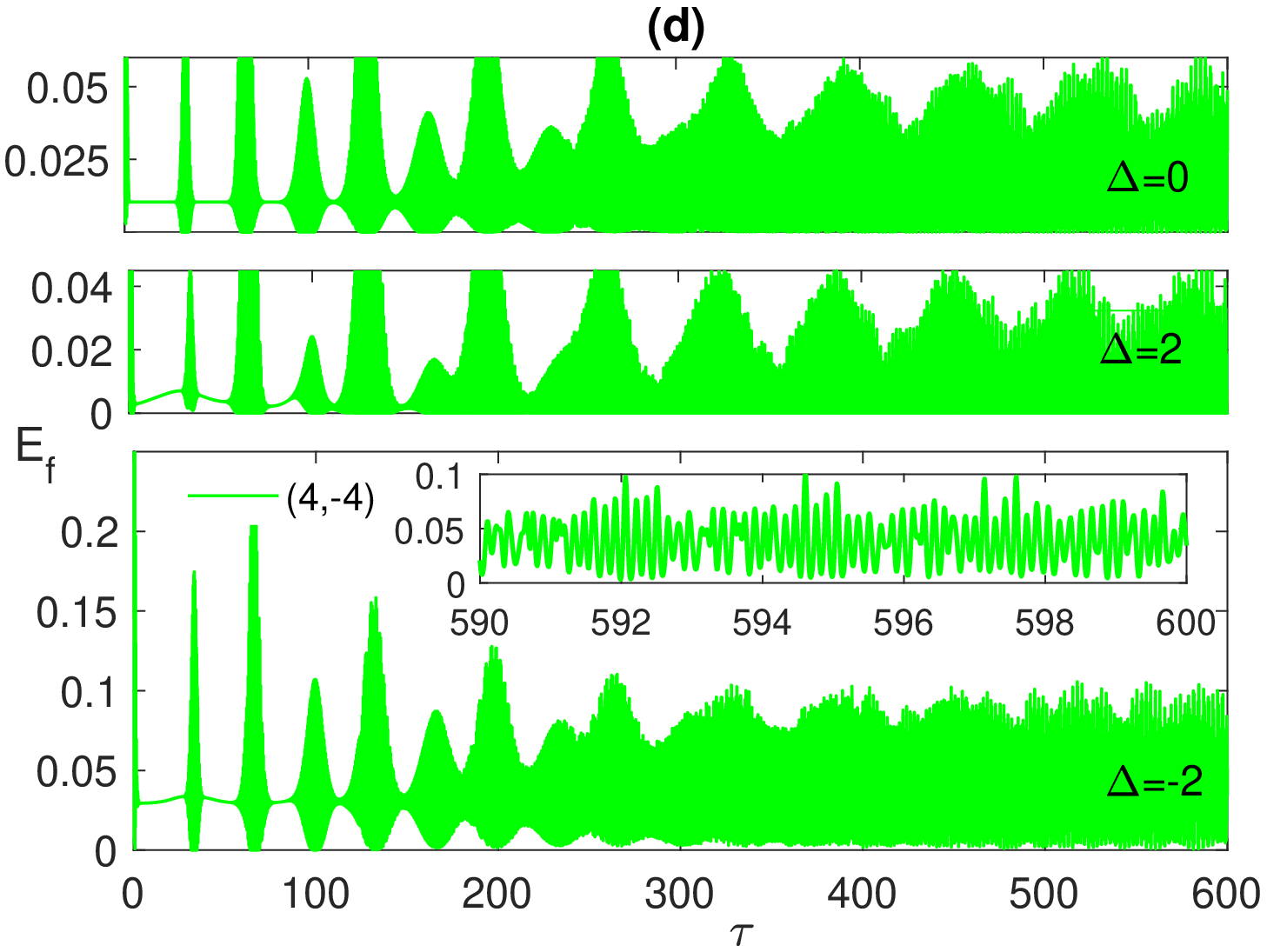}}\\
\caption{{
The entanglement $E_f$ versus the scaled time $\tau=\lambda_1 t$ with the two atoms are initially in a W-like state $\psi_{W}=(\vert g_{1}\rangle \vert g_{2}\rangle + \vert g_{1}\rangle \vert e_{2}\rangle + \vert e_{1}\rangle \vert g_{2}\rangle)/\sqrt{3}$ and the field is in a coherent state with $\bar{n}=100$, where in (a) $\Delta=0$; (b) $2$; (c) $-2$; (d) $0,2$ and $-2$. $(\lambda_2, J)$ take various values as shown in the legend in each panel.}}
\label{fig4}
\end{figure}

\begin{figure}[htbp]
 \centering
\subfigure{\includegraphics[width=12cm]{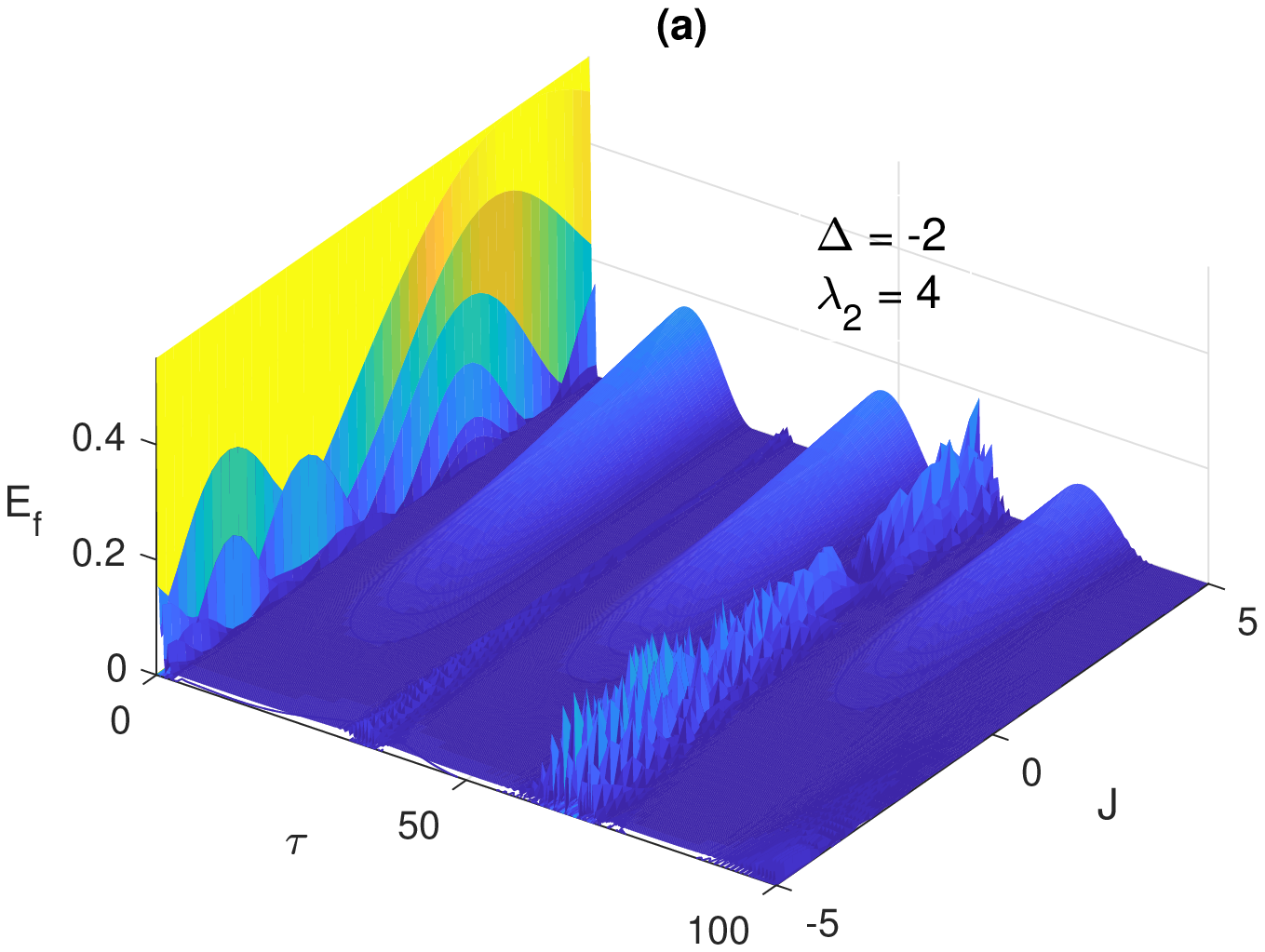}}\\\quad 
\subfigure{\includegraphics[width=10cm]{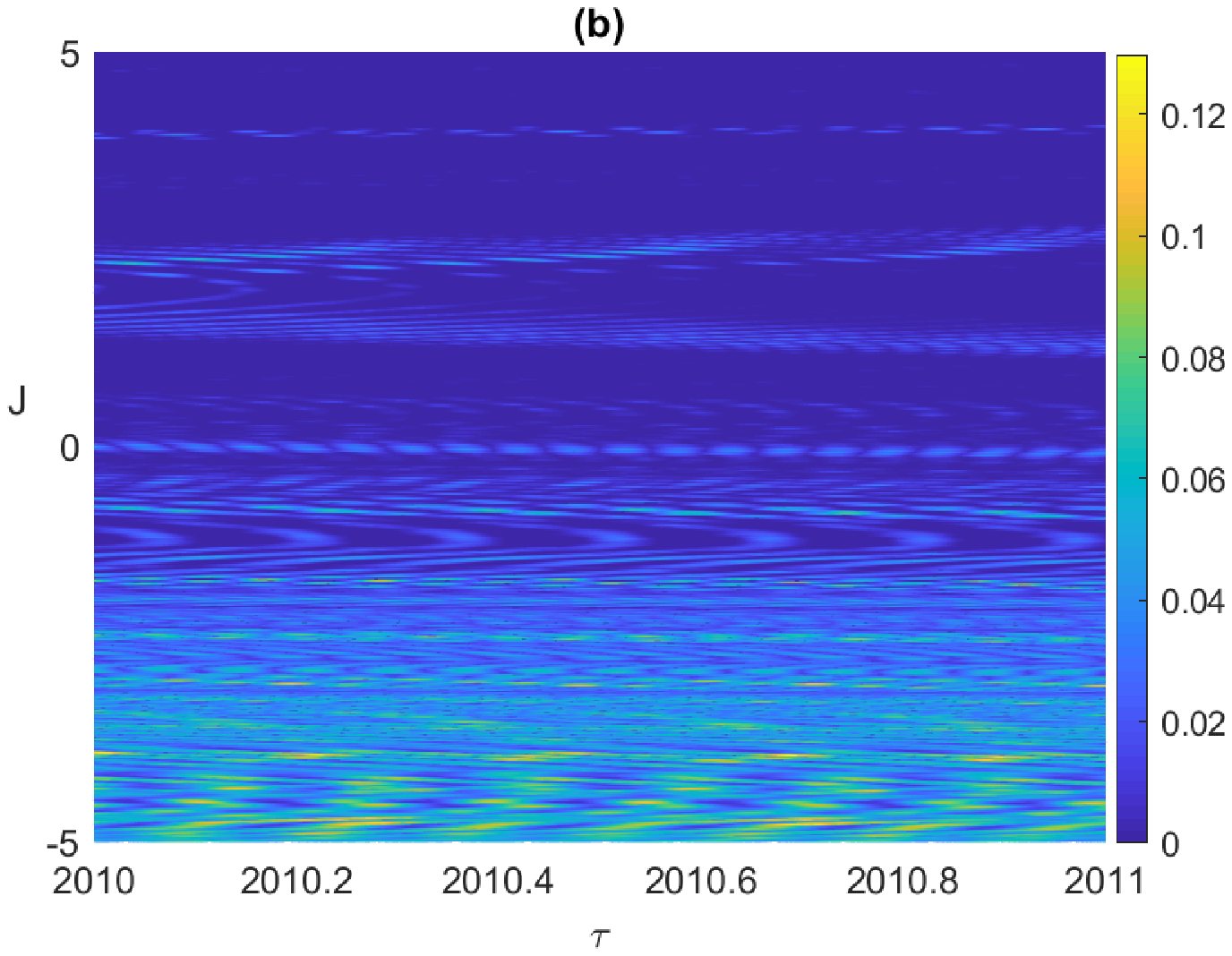}}\\ 
\caption{
The entanglement $E_f$ versus the scaled time $\tau=\lambda_1 t$ and the Ising coupling J
with the two atoms are initially in a W-like state $\psi_{W}$ and the field is in a coherent state where $\lambda_2=4$ and $\Delta=-2$, at early time in (a) and asymptotically at (b).}
\label{fig5}
\end{figure}

\begin{figure}[htbp]
 \centering
\subfigure{\includegraphics[width=6.4cm]{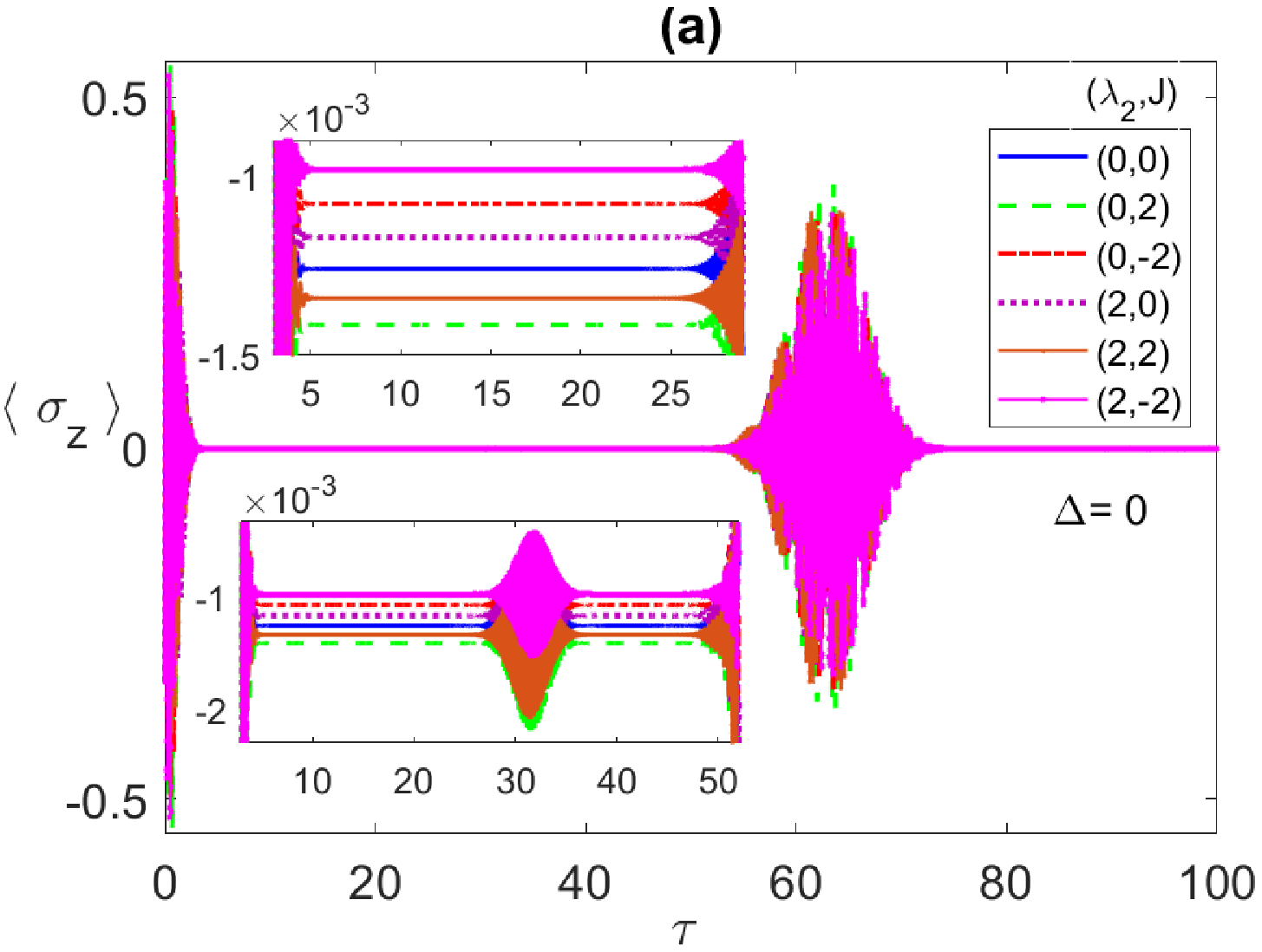}}\quad 
\subfigure{\includegraphics[width=6.4cm]{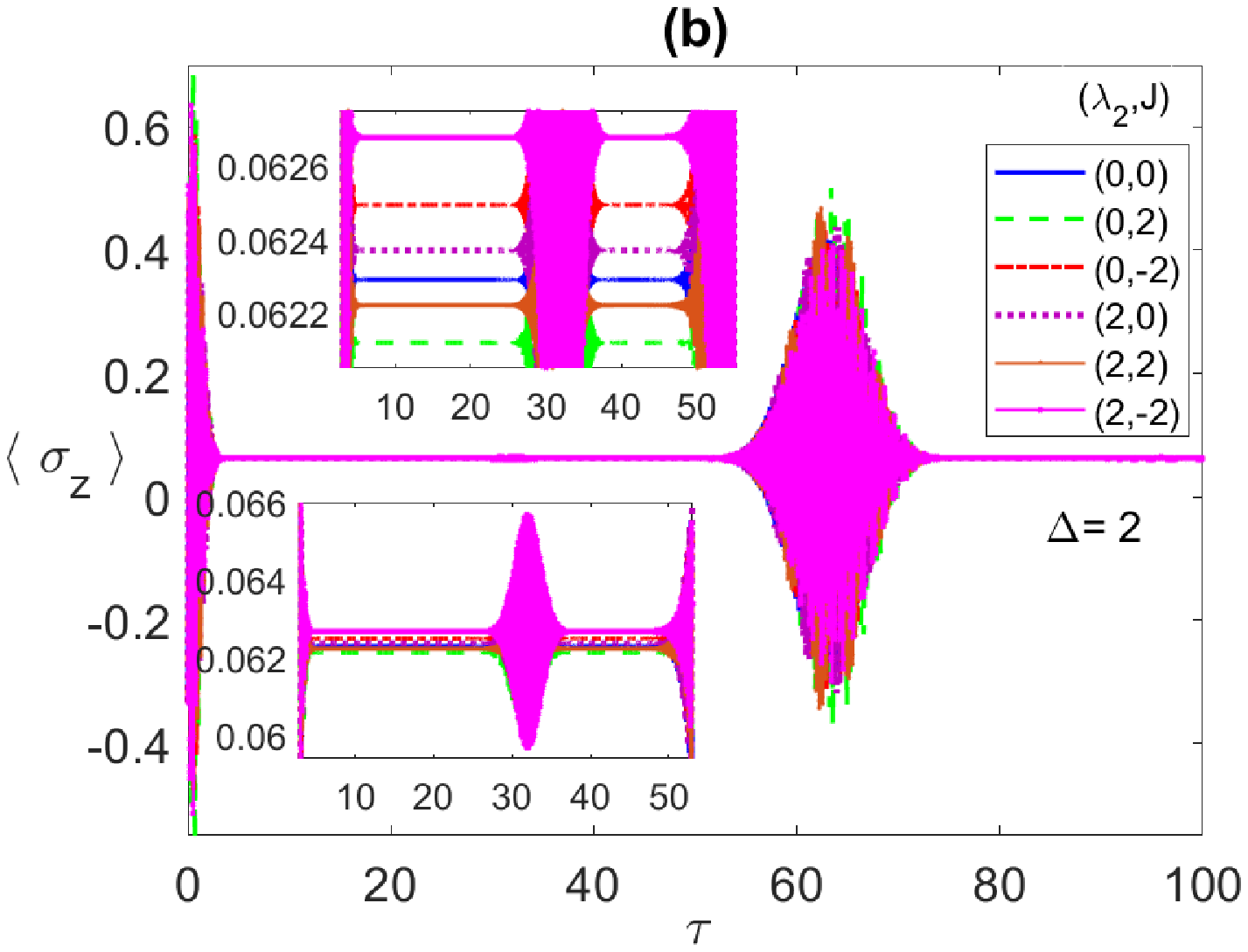}}\\
\subfigure{\includegraphics[width=6.4cm]{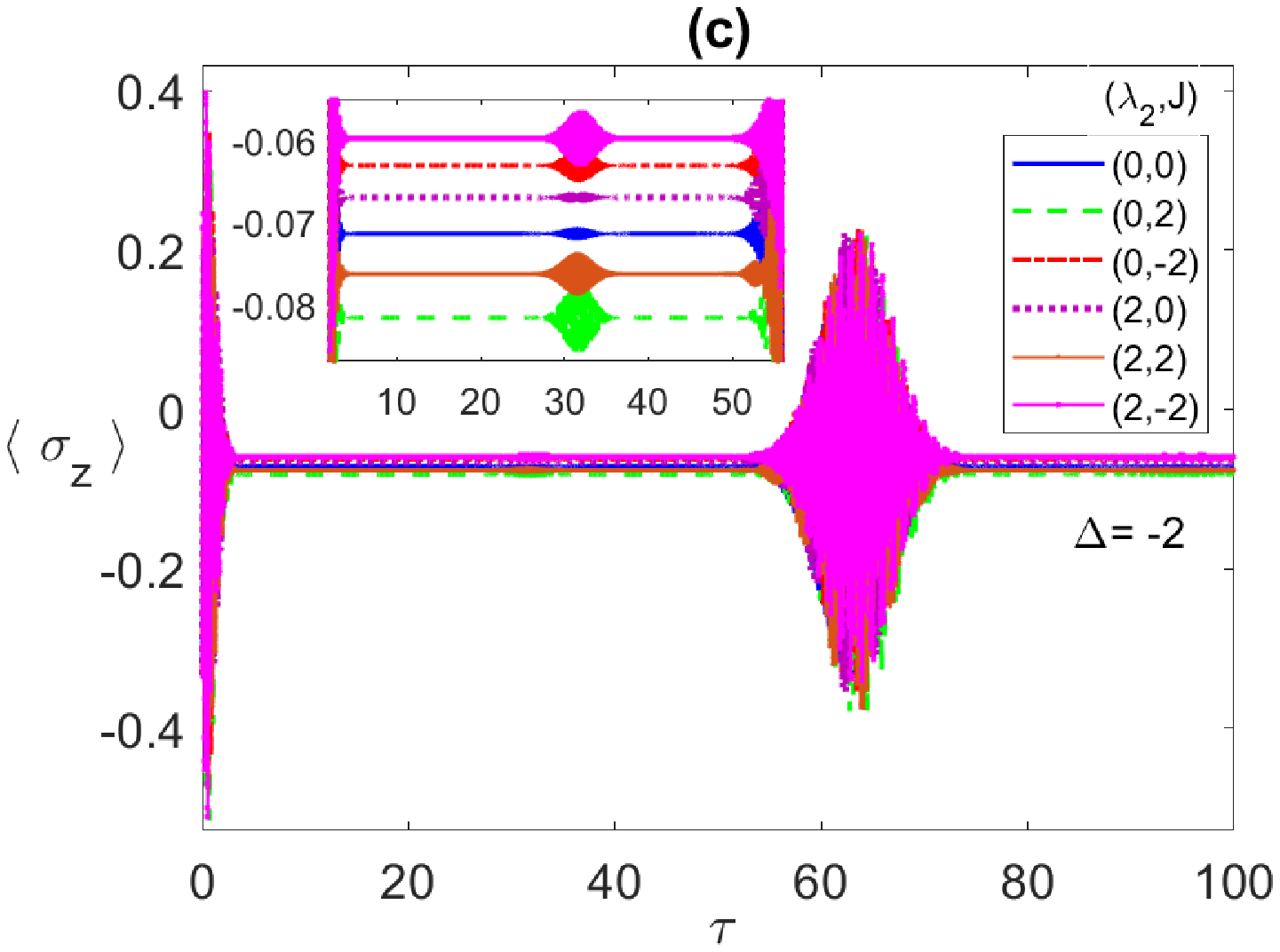}}\quad
\subfigure{\includegraphics[width=6.4cm]{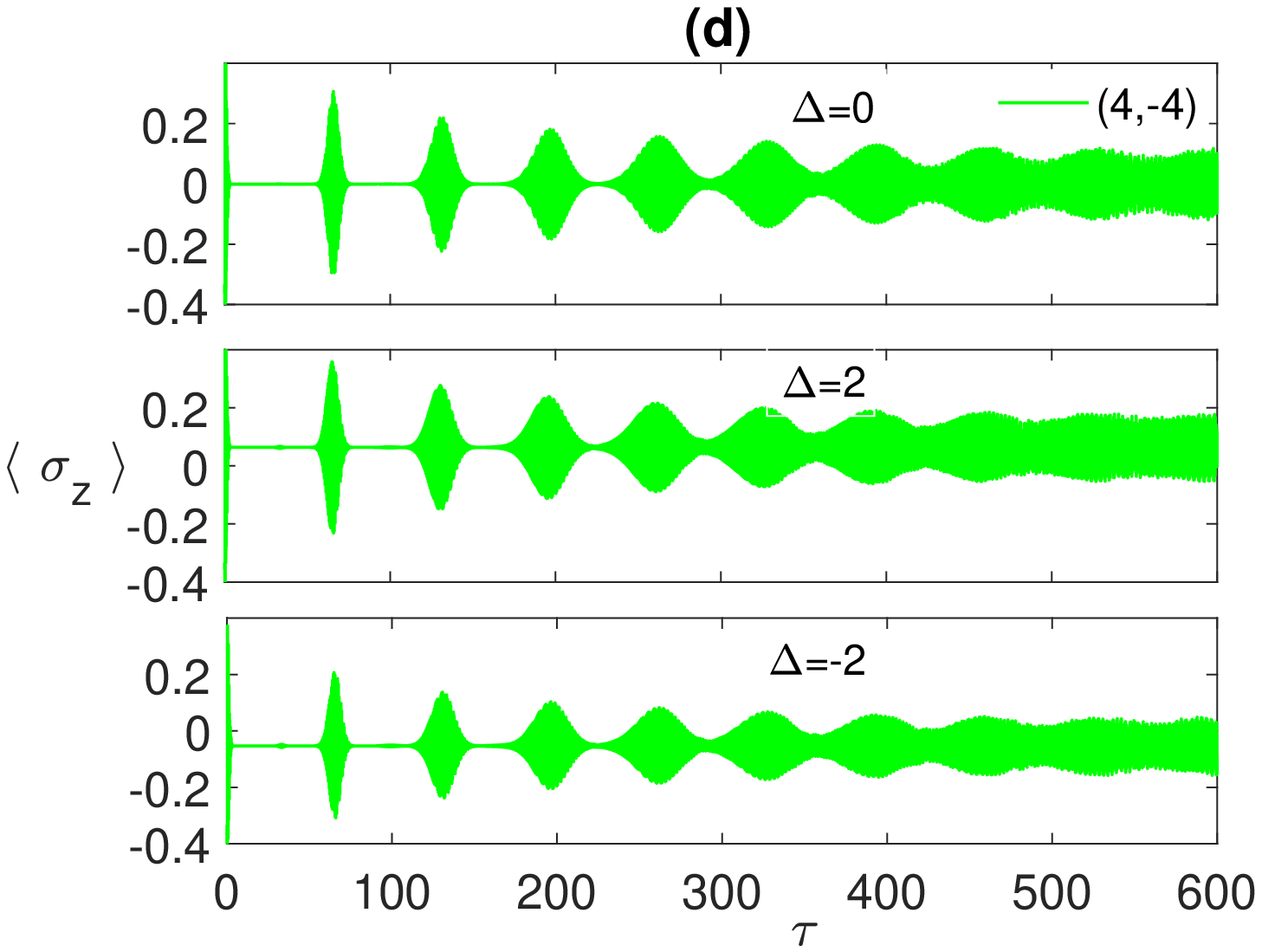}}\\
\caption{{
The population inversion $\langle \sigma_z \rangle$ versus the scaled time $\tau=\lambda_1 t$ with the two atoms are initially  in a W-like state $\psi_{W}=(\vert g_{1}\rangle \vert g_{2}\rangle + \vert g_{1}\rangle \vert e_{2}\rangle + \vert e_{1}\rangle \vert g_{2}\rangle)/\sqrt{3}$ and the field is in a coherent state  with $\bar{n}=100$, where in (a) $\Delta=0$; (b) $2$; (c) $-2$; (d) $0,2$ and $-2$. $(\lambda_2, J)$ take various values as shown in the legend in each panel.}}
\label{fig6}
\end{figure}

In Figs.~\ref{fig4}, \ref{fig5} and \ref{fig6}, we discuss the system dynamics starting from a partially entangled initial state that yields ESD upon evolution, namely, the W-like state $\psi_{W}=(\vert g_{1}\rangle \vert g_{2}\rangle + \vert g_{1}\rangle \vert e_{2}\rangle + \vert e_{1}\rangle \vert g_{2}\rangle)/\sqrt{3}$. The ESD periods that we used to observe in the maximally entangled initial states, for instance, the first period from $\tau \sim 3$ to $28$, are interrupted, in the current case, in the middle by a single reviving peak at all choices of $\lambda_2$ and $J$, as shown in all panels of Fig.~\ref{fig4}.
The width of these peaks and their heights vary depending on the different system parameters. The maximum height corresponds to (0,2), i.e. with only AF Ising coupling between the two atoms, at all $\Delta$ values, while the largest width takes place at $\Delta=2$ as shown in Figs.~\ref{fig4}(b). The ESD can be removed partially, particularly in the first ESD period at the detuning $\Delta=-2$, mostly by combining the dipole and the FM Ising couplings as shown in the inner inset of Fig.~\ref{fig4}(c). Interestingly, increasing the combined values of the dipole and FM Ising interactions further, (4,-4), as shown in Fig.~\ref{fig4}(d), yields a complete elimination of the ESD at the different detuning values, most effectively at $\Delta=-2$ and less effectively at $\Delta= 2$. The asymptotic behavior of the entanglement, illustrated in Fig.~\ref{fig4}(d), shows that the entanglement collapse periods diminish with time until completely disappearing and the entanglement takes an oscillatory form without any ESD interruptions as illustrated in the inset in the bottom panel in Fig.~\ref{fig4}(d). 
In Fig.~\ref{fig5}(a), we plot the entanglement dynamics versus the Ising coupling $J$ over a wide range form $-5$ to $5$. As can be noticed, applying high values of Ferromagnetic Ising coupling eliminates the ESD, while having an anti-Ferromagnetic coupling leads to ESD, although it is accompanied by revivals, as shown in the figure. The same effect of the Ising coupling type and strength is confirmed in Fig.~\ref{fig5}(b), where the contour plot of the asymptotic entanglement is presented.

The atomic population is illustrated in Fig.~\ref{fig6}(a), (b) and (c) for the detuning values $0$, $2$ and $-2$ respectively. Again, the separation of the collapse lines from the (0,0) blue line, at all parameter values match the same one of the lines in the entanglement case, Fig.~\ref{fig4}. The collapse lines separation from each other increases from $\Delta=0$ to $2$ and mostly at $-2$. The detuning effect is clear in shifting the collapse lines upwards, for $\Delta=2$, and downwards, for $\Delta=-2$, compared with the $\Delta=0$ case. Noticeably, the inverted reflection symmetry of the population lines, as we turn from $\Delta=2$ to $\Delta=-2$, that was observed in the Bell state case, in Fig.~\ref{fig3}, is not seen in the current case. In Fig.~\ref{fig6}(d), we depict the asymptotic dynamics of the population inversion at the parameters choice (4,-4), at the detuning values 0, 2 and -2 respectively in the top, middle and bottom panels respectively. As can be seen, in all cases, the collapse periods shrinks until completely disappearing, where the dynamics profile becomes a continuous irregular oscillation, which resembles the entanglement function asymptotic behavior and provides an explanation for it, where the exchange of energy between the two atoms and the field becomes continuous with no interruption, enhancing the mediated entanglement between the two atoms.
\subsection{Disentangled initial state}
Now we turn to another different type of initial state, which is completely separable (disentangled), where the state of the two atoms is a linear combination of all the basis states, namely  $\psi_{L}=(\vert g_{1}\rangle \vert g_{2}\rangle + \vert g_{1}\rangle \vert e_{2}\rangle + \vert e_{1}\rangle \vert g_{2}\rangle + \vert e_{1}\rangle \vert e_{2}\rangle)/\sqrt{4}$, while the radiation field is in a coherent state as previously described.
\begin{figure}[htbp]
 \centering
\subfigure{\includegraphics[width=6.4cm]{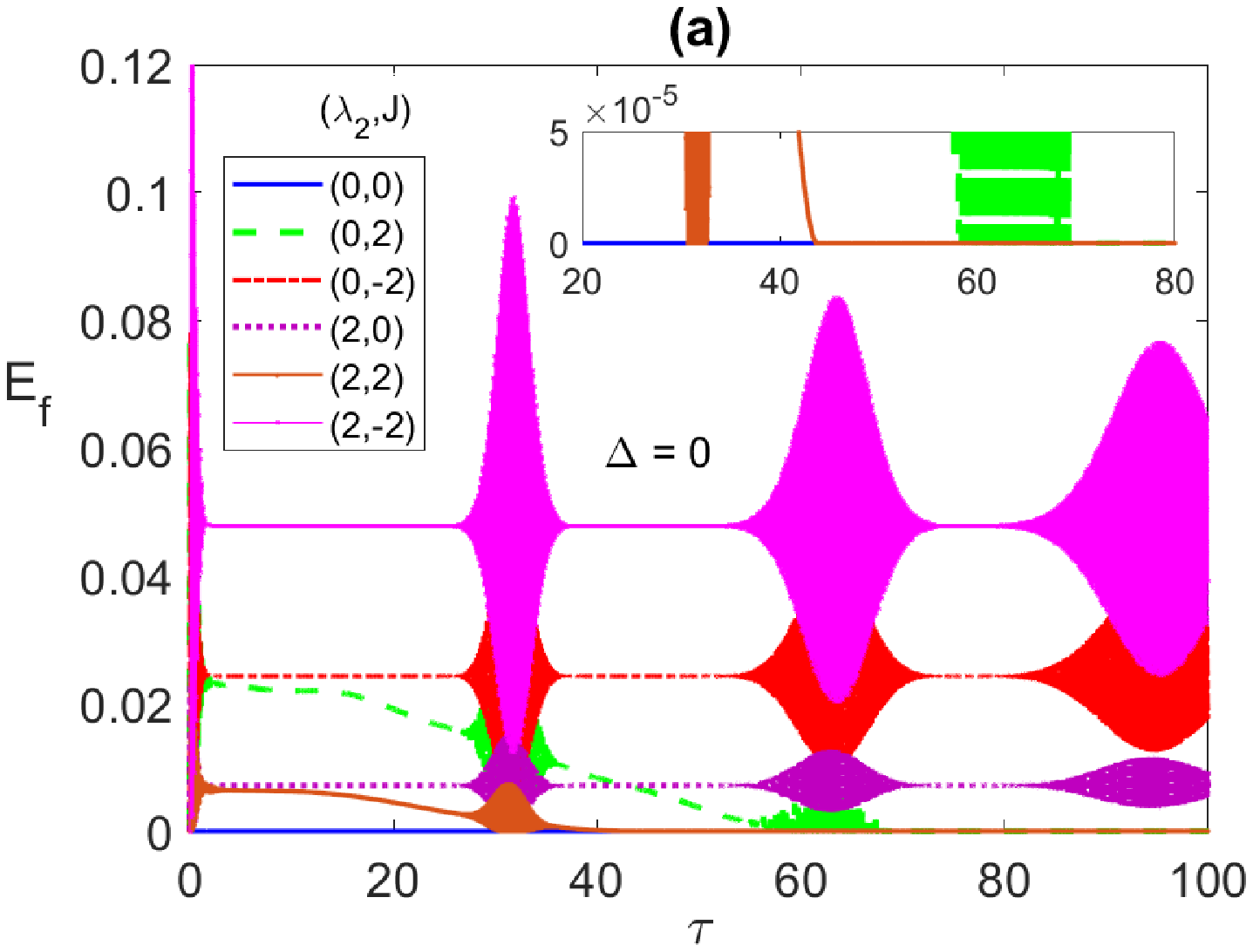}}\quad 
\subfigure{\includegraphics[width=6.4cm]{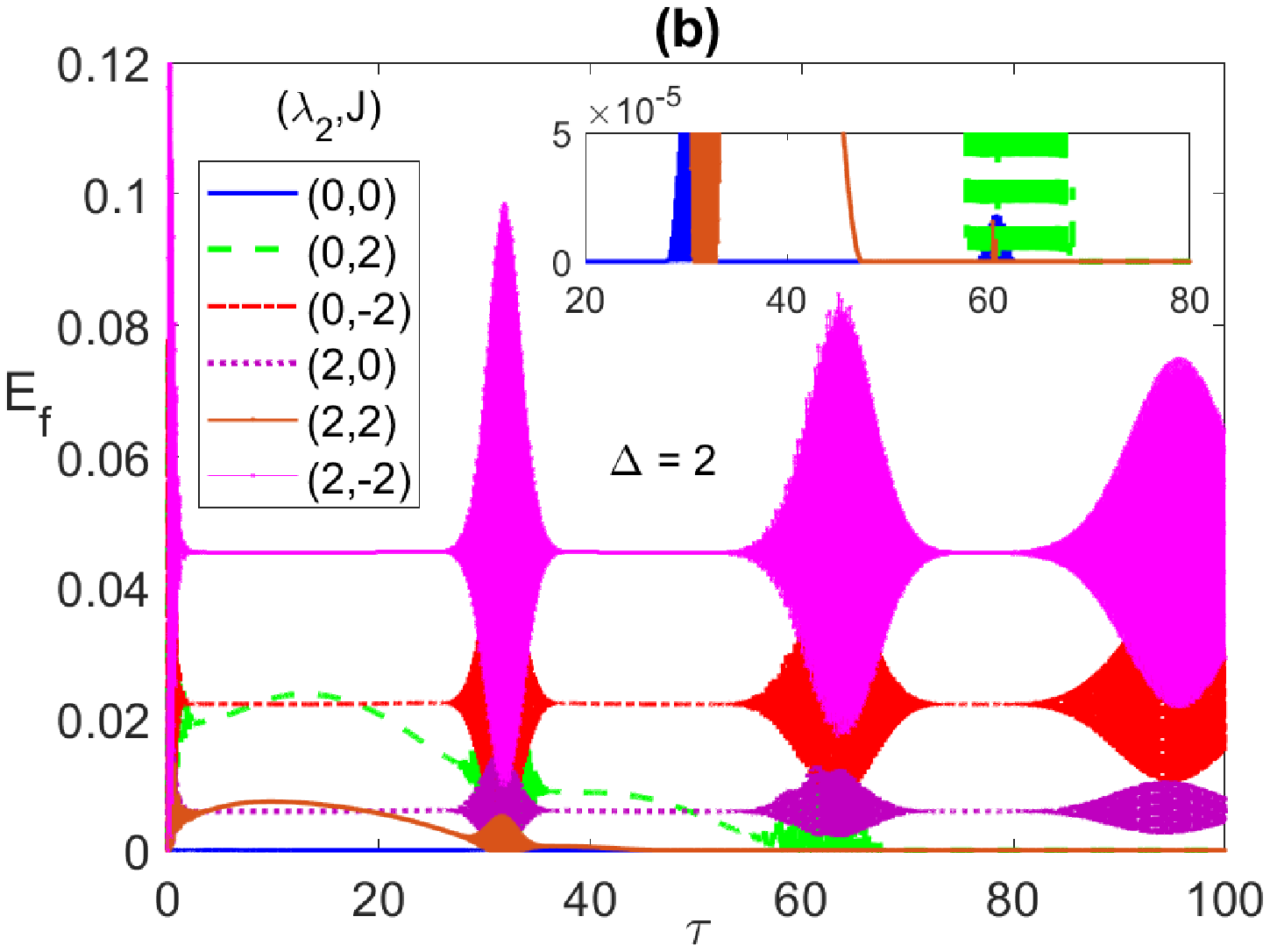}}\\
\subfigure{\includegraphics[width=6.4cm]{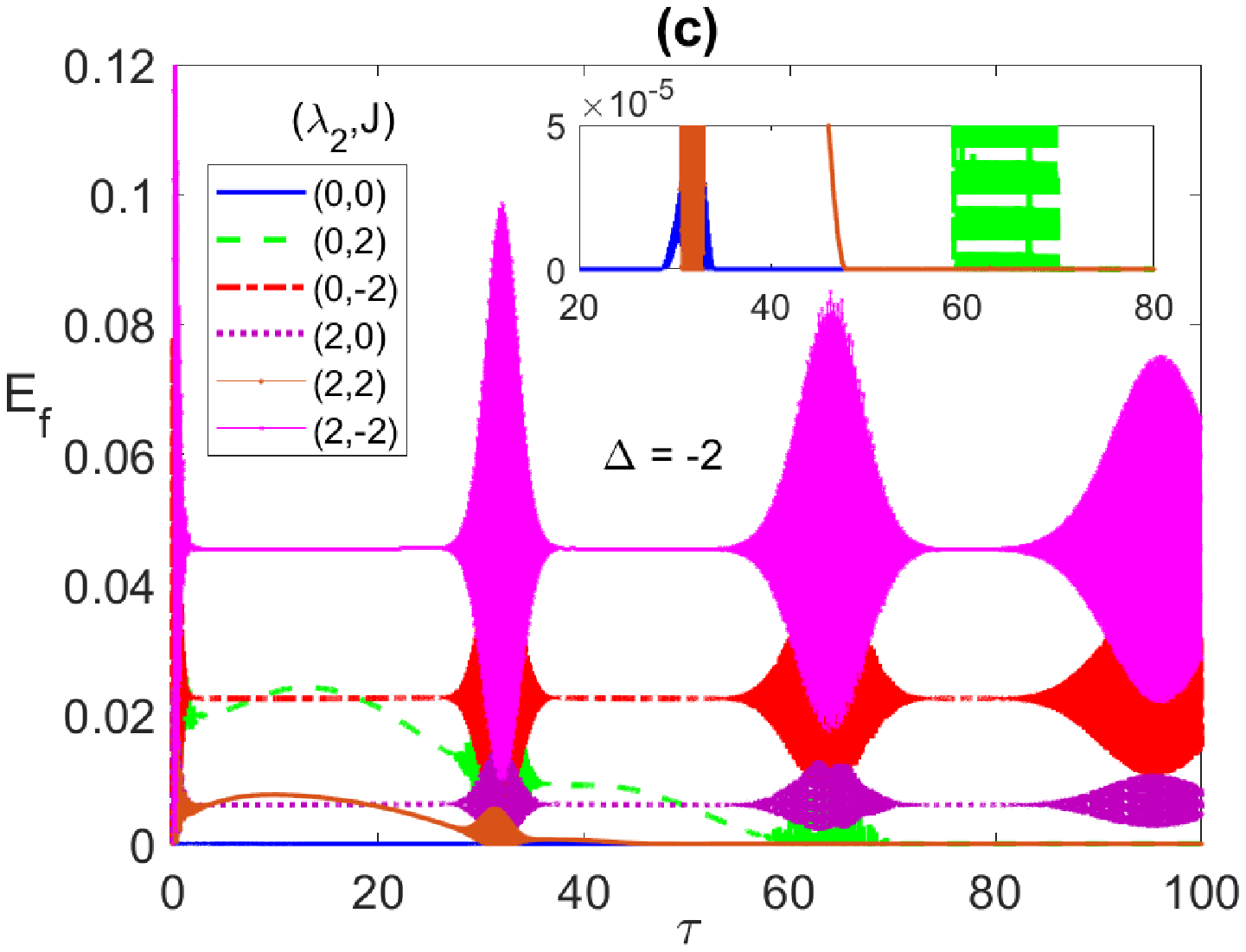}}\quad 
\subfigure{\includegraphics[width=6.4cm]{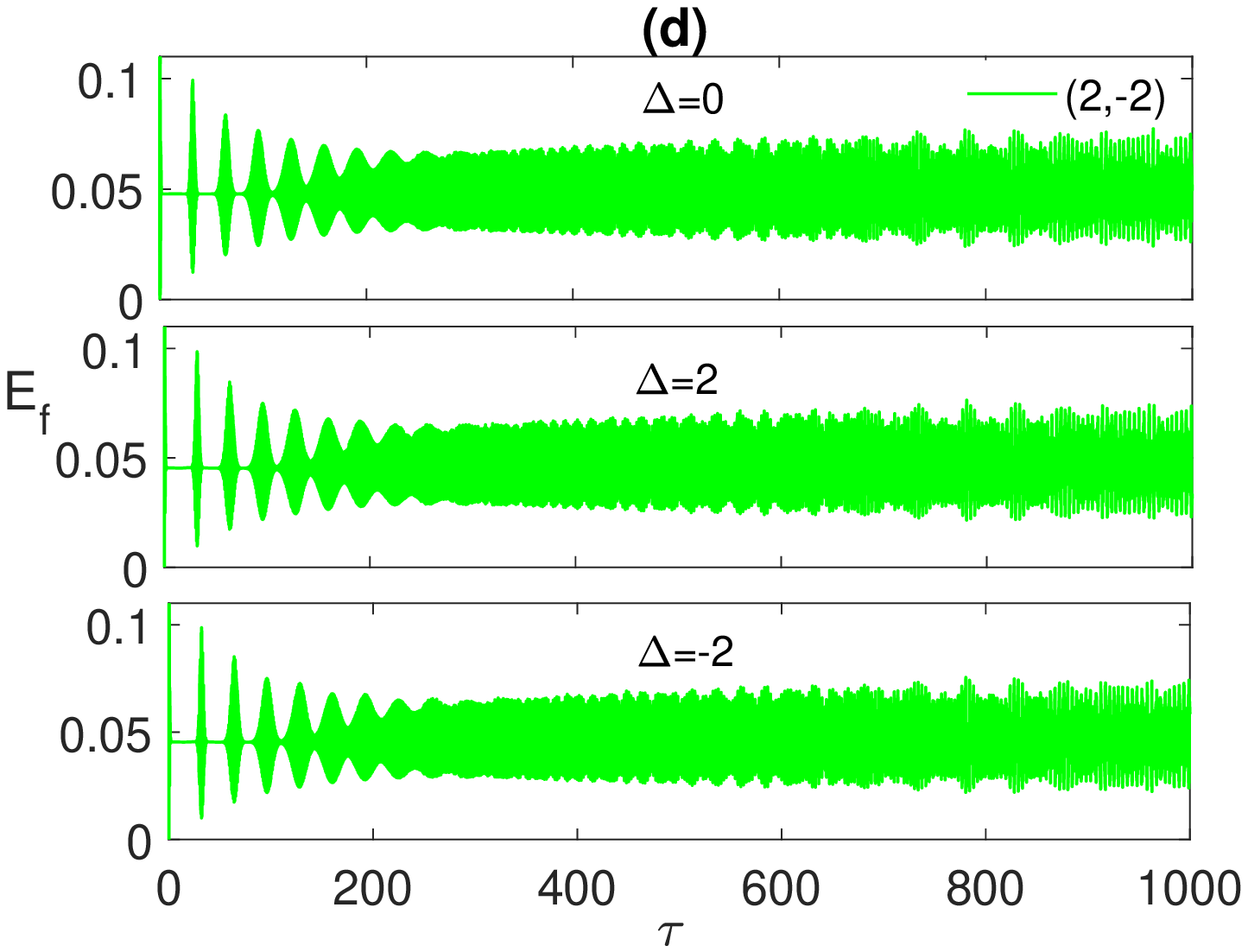}}\\
\caption{{
The entanglement $E_f$ versus the scaled time $\tau=\lambda_1 t$ with the two atoms are initially in a separable state $\psi_{L}=(\vert g_{1}\rangle \vert g_{2}\rangle + \vert g_{1}\rangle \vert e_{2}\rangle + \vert e_{1}\rangle \vert g_{2}\rangle + \vert e_{1}\rangle \vert e_{2}\rangle)/\sqrt{4}$ and the field is in a coherent state with $\bar{n}=100$, where in (a) $\Delta=0$; (b) $2$; (c) $-2$; (d) $0,2$ and $-2$. $(\lambda_2, J)$ take various values as shown in the legend in each panel.}}
\label{fig7}
\end{figure}
\begin{figure}[htbp]
 \centering
\subfigure{\includegraphics[width=11cm]{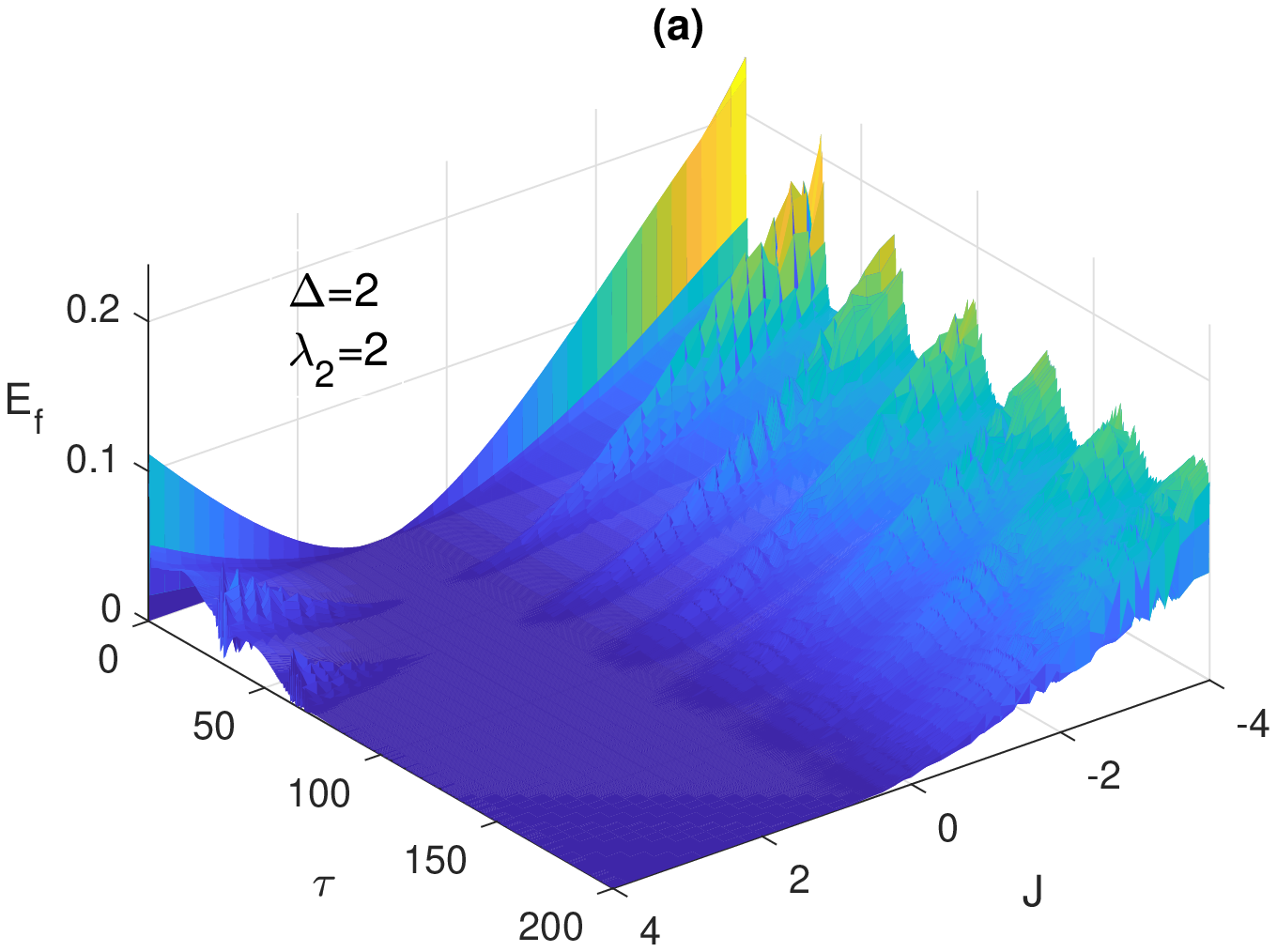}}\quad 
\subfigure{\includegraphics[width=11cm]{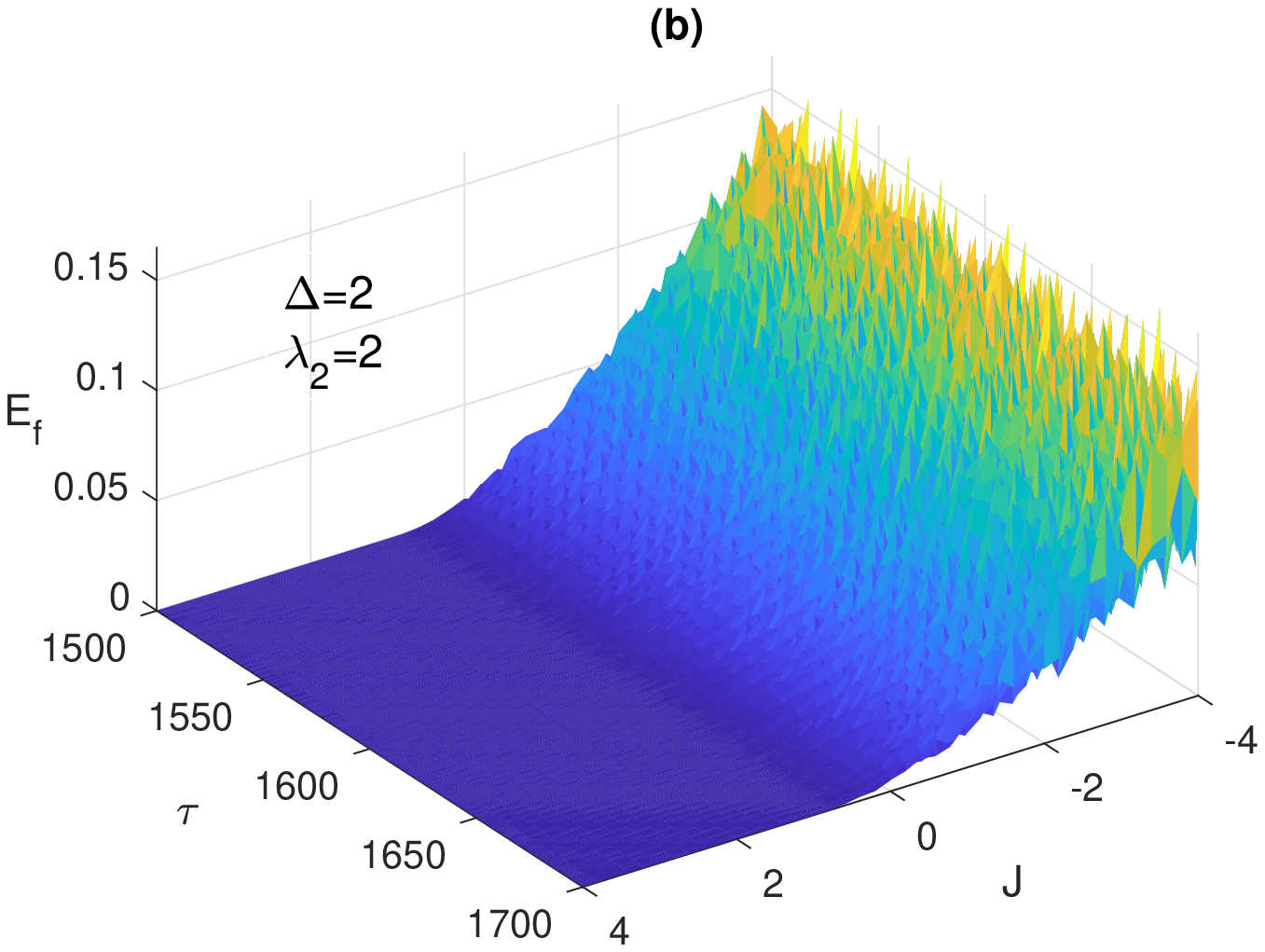}}\\
\caption{
The entanglement $E_f$ versus the scaled time $\tau=\lambda_1 t$ and the Ising coupling J
with the two atoms are initially in the separable state $\psi_{L}$ and the field is in a coherent state, where $\lambda_2=2$ and $\Delta=2$, at early time in (a) and asymptotically at (b).}
\label{fig8}
\end{figure}

\begin{figure}[htbp]
\centering
\subfigure{\includegraphics[width=6.4cm]{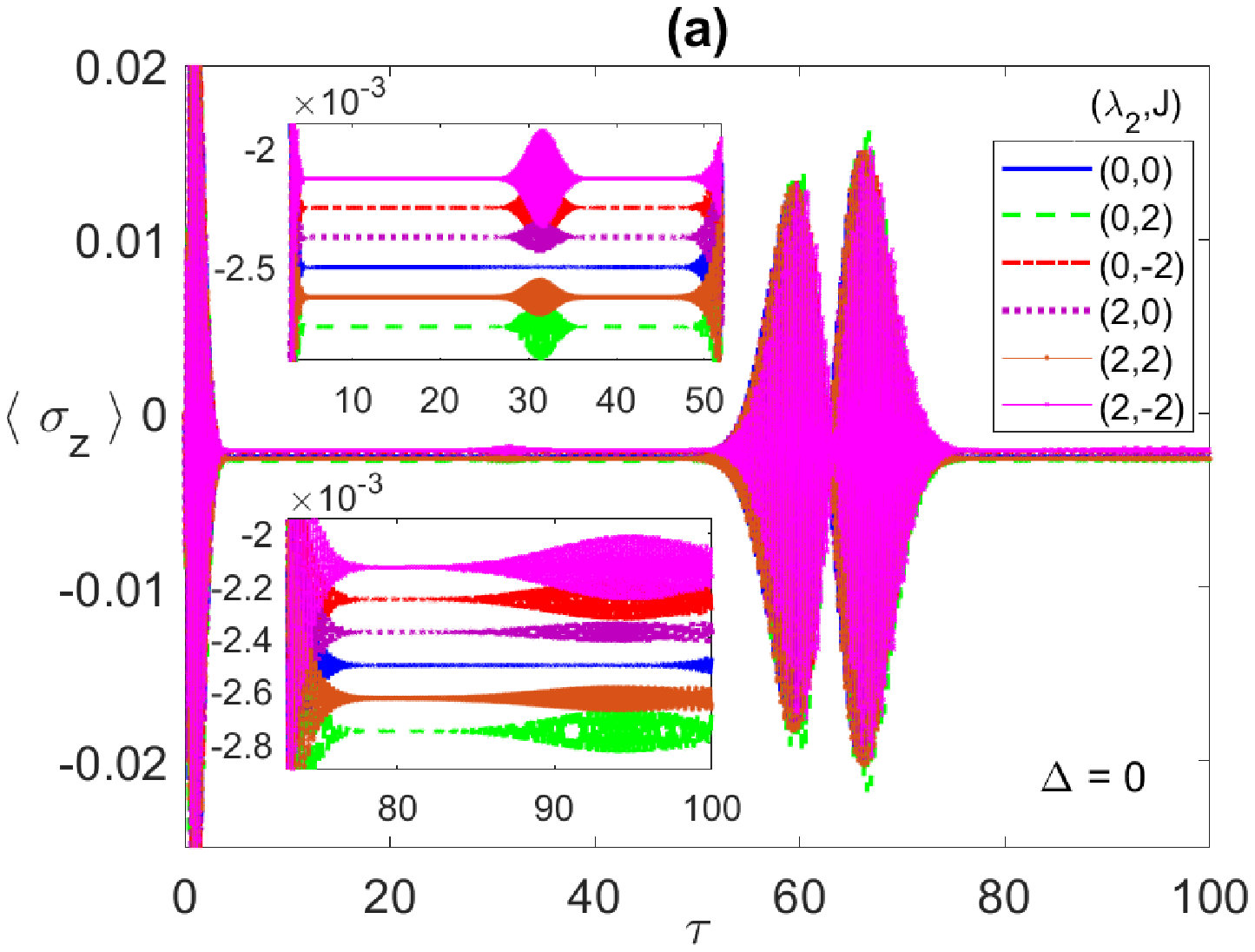}}\quad 
\subfigure{\includegraphics[width=6.4cm]{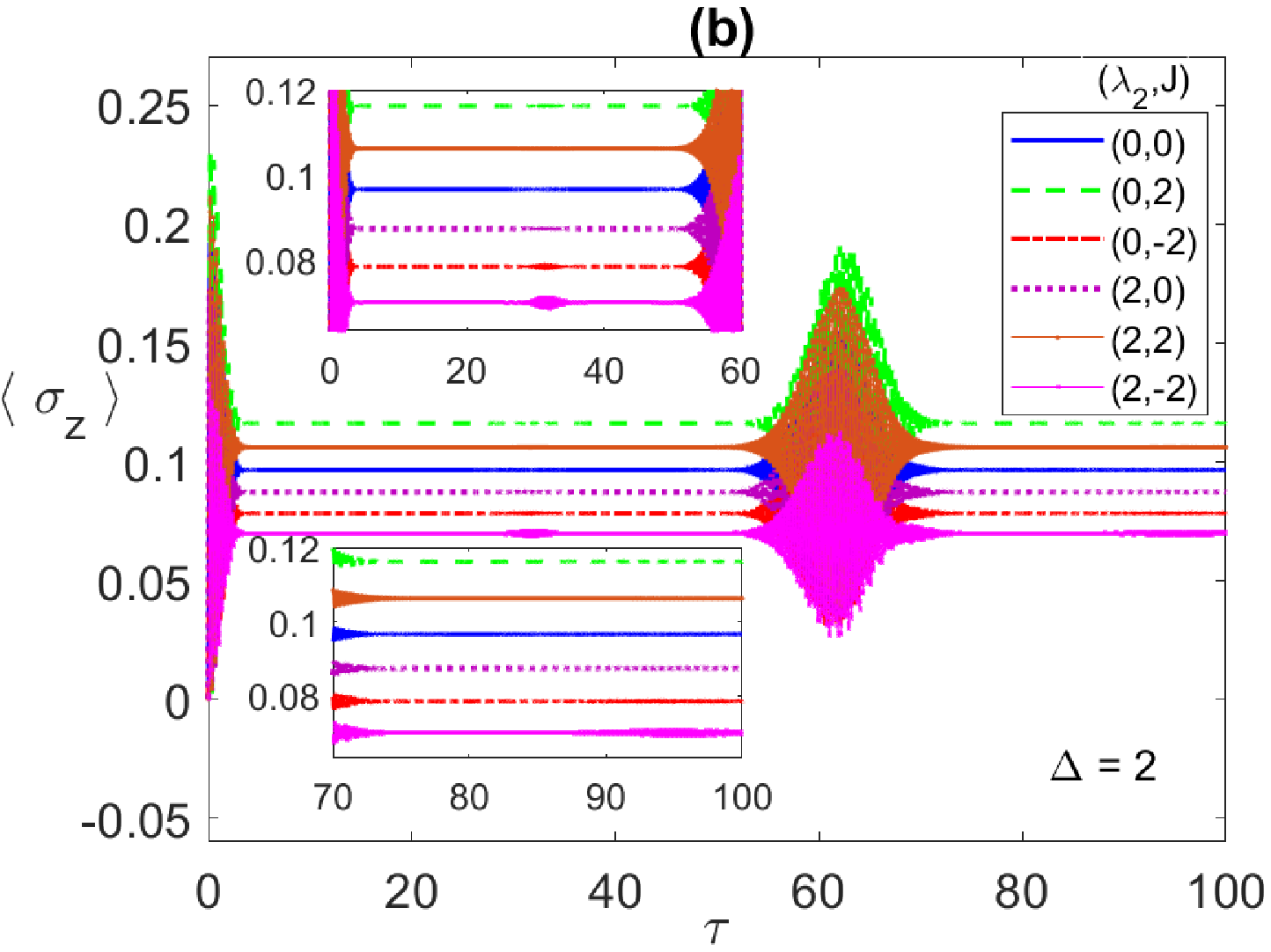}}\\
\subfigure{\includegraphics[width=6.4cm]{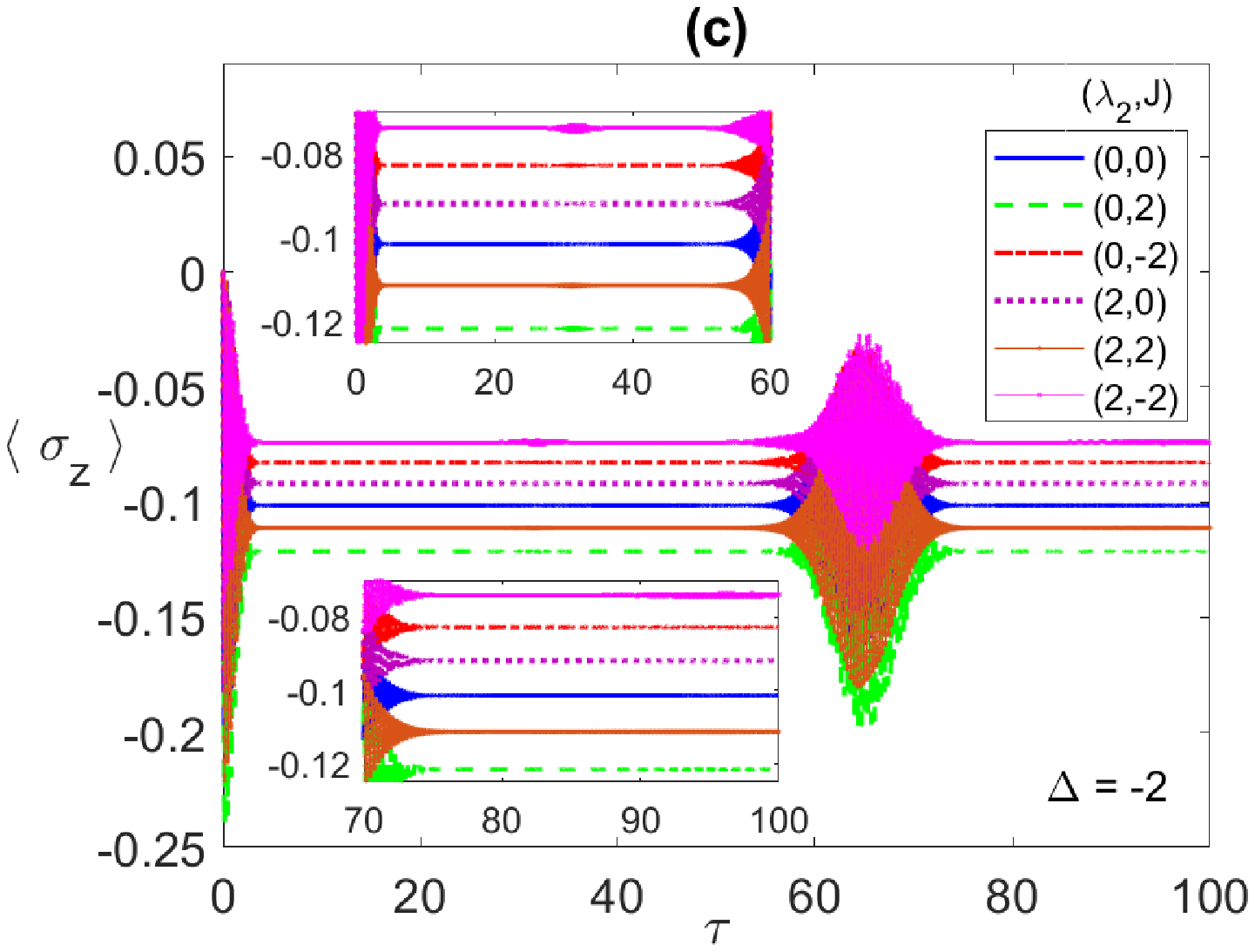}}\quad 
\subfigure{\includegraphics[width=6.4cm]{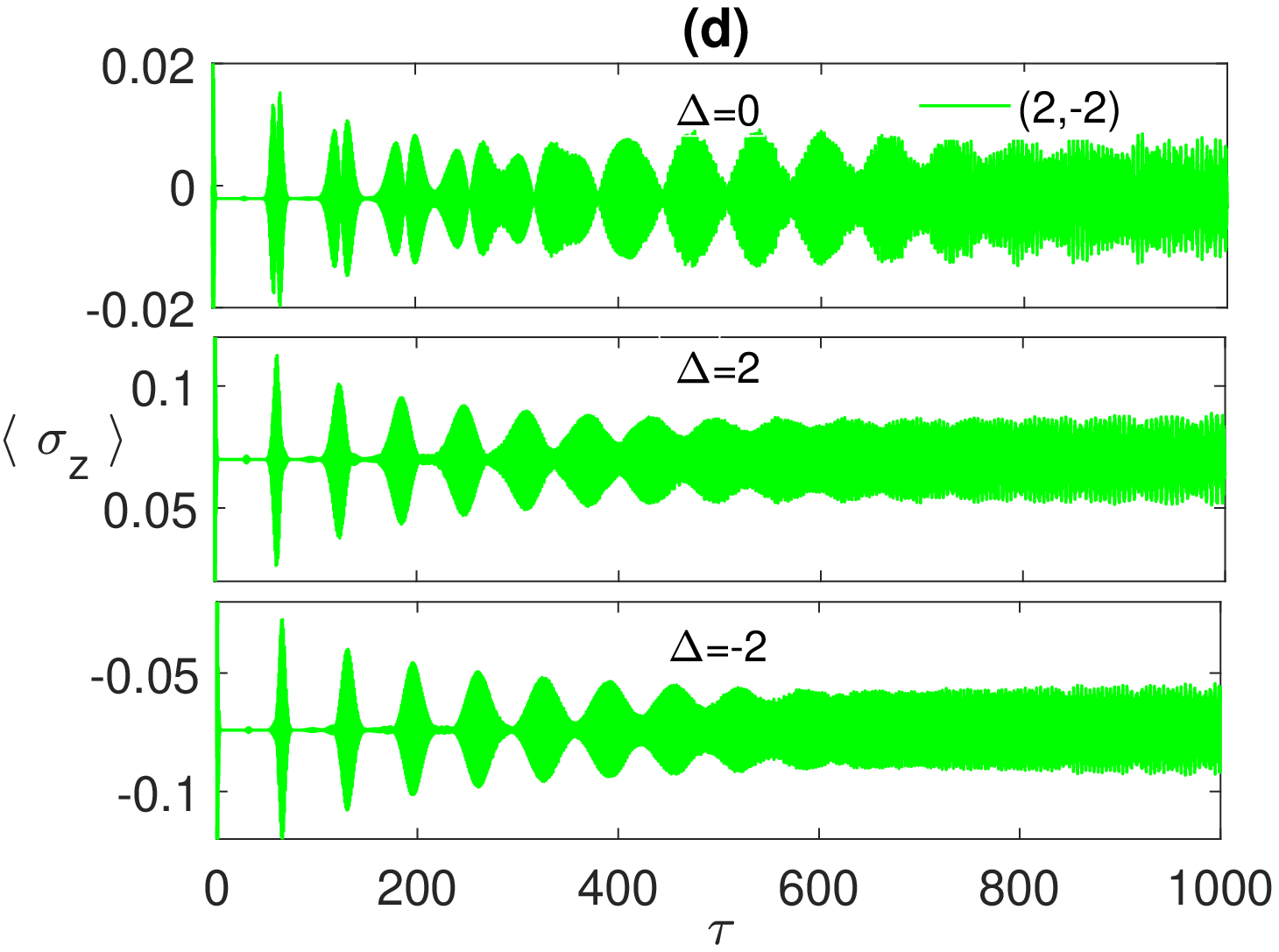}}\\
\caption{
The population inversion $\langle \sigma_z \rangle$ versus the scaled time $\tau=\lambda_1 t$ with the two atoms are initially in a separable state $\psi_{L}=(\vert g_{1}\rangle \vert g_{2}\rangle + \vert g_{1}\rangle \vert e_{2}\rangle + \vert e_{1}\rangle \vert g_{2}\rangle + \vert e_{1}\rangle \vert e_{2}\rangle)/\sqrt{4}$ and the field is in a coherent state with $\bar{n}=100$, where in (a) $\Delta=0$; (b) $2$; (c) $-2$; (d) $0,2$ and $-2$. $(\lambda_2, J)$ take various values as shown in the legend in each panel.}
\label{fig9}
\end{figure}

This interesting initial state maintains zero entanglement between the two non-interacting atoms at zero detuning forever, as shown in Fig.~\ref{fig7}(a) (blue line). Turning on a low AF Ising coupling, (0,2), causes the entanglement to rise  initially from zero reaching a low peak before eventually suffering ESD, without any future revival (Green line). However, applying a low FM Ising coupling, (0,-2), saves the entanglement from any sudden death and makes it maintain a non-zero value for the whole time, with a collapse-revival-like pattern (red line). Also, turning on the dipole interaction in the absence of the Ising interaction, (2,0), causes the entanglement to exhibit a very similar behavior to that of 
the setup (0,-2) but at a smaller mean value (purple line). Adding AF Ising interaction to the dipole one, (2,2), turns out to be devastating to the entanglement, where the ESD is reached early right after the entanglement rises from zero (brown line), as illustrated in the inner inset. However, as we have observed in the previous cases, combining the dipole and FM Ising interactions is very effective in removing the ESD and boosting the entanglement, which is the case here too, where applying (2,-2) causes the entanglement to rise from zero to a high persisting mean value (magenta line), compared with the other choices. Remarkably, changing the detuning to a non-zero value $(\pm 2)$ does not show a significant impact on the entanglement dynamics, as can be noticed in Figs.~\ref{fig7}(b) and (c), except for quite small reviving peaks in the case (0,0) before the entanglement eventually vanishes again completely once and for all. The negligible effect of changing the detuning value can be seen also in Fig.~\ref{fig7}(d), where the asymptotic behavior of the entanglement shows the disappearance of the collapse periods and the emergence of a continuous irregular oscillation with a very similar profile at all the detuning values. 
Fig.~\ref{fig8} shows the decisive role the Ising coupling plays when the system starts from the separable initial state $\psi_{L}$, where the entanglement dynamics is tested versus the Ising coupling $J$. The ESD takes place at early times and asymptotically for positive values of J that are higher than approximately $0.95$, but can be completely eliminated for smaller positive values of J and all negative values.

In fact, the critical role played by the Ising coupling in controlling the system dynamics and asymptotic behavior is highlighted when one particularly compares the 3-dimensional plots in Figs.~\ref{fig2}, \ref{fig5} and \ref{fig8}. They clearly show how the type of Ising coupling (Ferromagnetic or anti-Ferromagnetic) and its strength, depending on the initial state, may lead to completely different entanglement profiles. While Fig.~\ref{fig2}(b) illustrates how a strong Ising coupling (of any type) could be devastating for the asymptotic entanglement starting from an initial correlated Bell state, Fig.~\ref{fig2}(c) shows how the same coupling can considerably boost the asymptotic entanglement starting form an anti-correlated Bell state. On the other hand, Fig.~\ref{fig5} shows how the effect of the Ising coupling type is crucial when the system starts from a partially entangled (W$-$like) state, where only a Ferromagnetic coupling is capable of eliminating the ESD at early times and asymptotically as illustrated in Fig.~\ref{fig5}(a) and (b) respectively. Interestingly, when the system starts from the separable state $\psi_{L}$, the Ising coupling parameter has a very peculiar impact, it can be used as a switch that turns on and off the entanglement asymptotically by crossing a critical value $J_c$, where the entanglement vanishes for coupling values higher than $J_c$, as illustrated in Fig.~\ref{fig8}.

The behavior of the population inversion differs significantly as the detuning parameter is varied, as shown in Fig.~\ref{fig9}. The zero detuning case is illustrated in Fig.~\ref{fig9}(a), which exhibits a much smaller oscillation amplitudes compared with the non-zero detuning cases. The non-zero detuning, positive or negative, causes the collapse lines to spread over a wider range compared with the zero detuning case, as illustrated in Fig.~\ref{fig9}(a) and (b). The positive detuning shifts the collapse line upwards, whereas the negative one shifts the lines downwards compared with the zero detuning case. Remarkably, similar to the maximally entangled initial state case and in contrary to the w-state one, the collapse-revival lines in the non-detuning cases are inverted images from each other. The asymptotic behavior of the atomic population, at (2,-2), is depicted in Fig.~\ref{fig9}(d) for the three different detuning cases. The profiles of the positive and negative detunings are very similar, as shown in the middle and bottom panels of Fig.~\ref{fig9}(d), though the main values are positive and negative respectively. The zero detuning profile, which looks different from the non-zero detuning, still shows a collapse-revival behavior, as shown in the upper panel of Fig.~\ref{fig9}(d). In all cases, the collapse periods disappear asymptotically and the profile turns to an oscillatory one to synchronize with the entanglement asymptotic dynamics. The separable initial states $\vert e_{1}\rangle \vert e_{2}\rangle$ and $\vert g_{1}\rangle \vert g_{2}\rangle$ do not show ESD upon evolution and therefore are not discussed here.

\begin{figure}[htbp]
 \centering
\subfigure{\includegraphics[width=6.4cm]{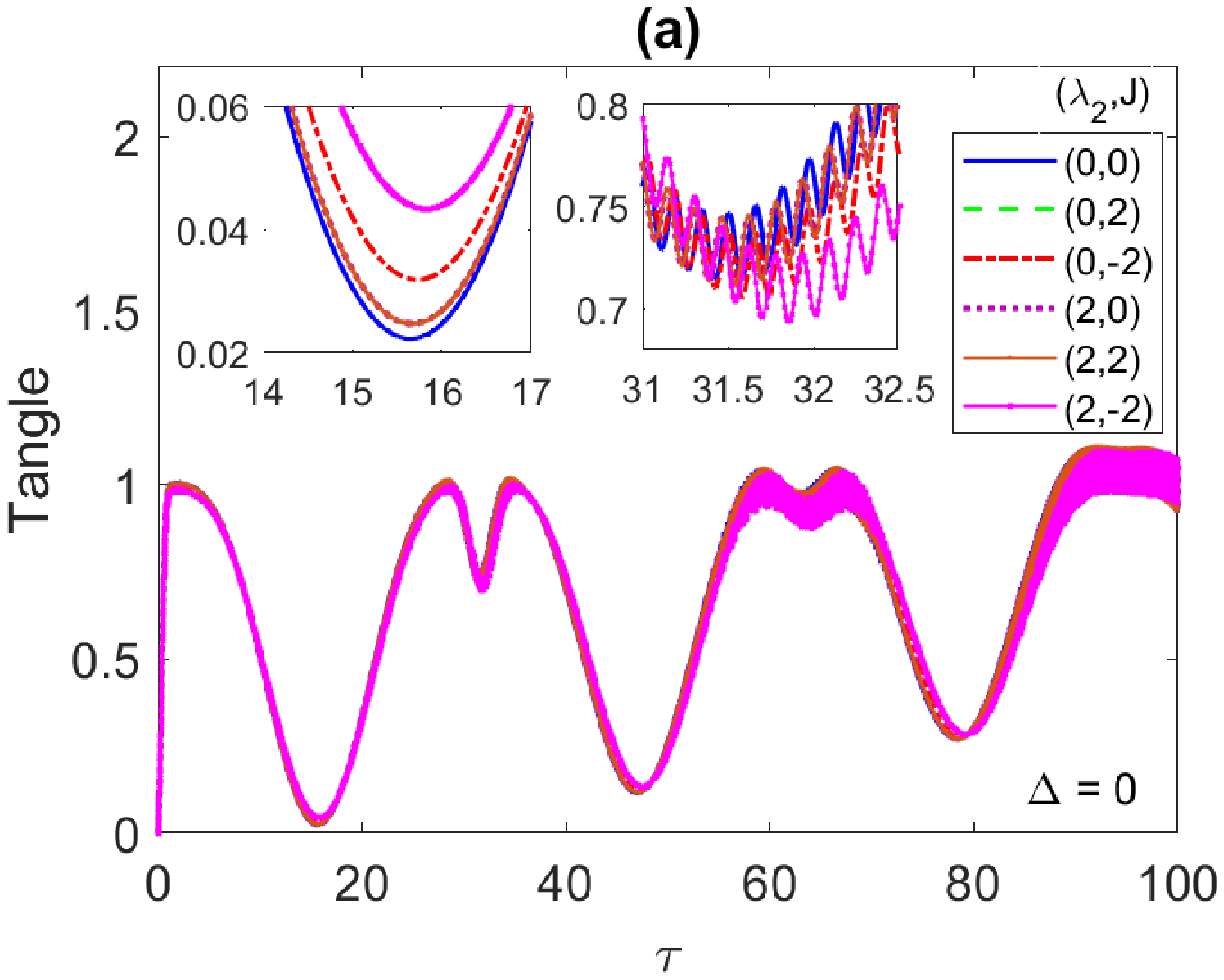}}\quad 
\subfigure{\includegraphics[width=6.4cm]{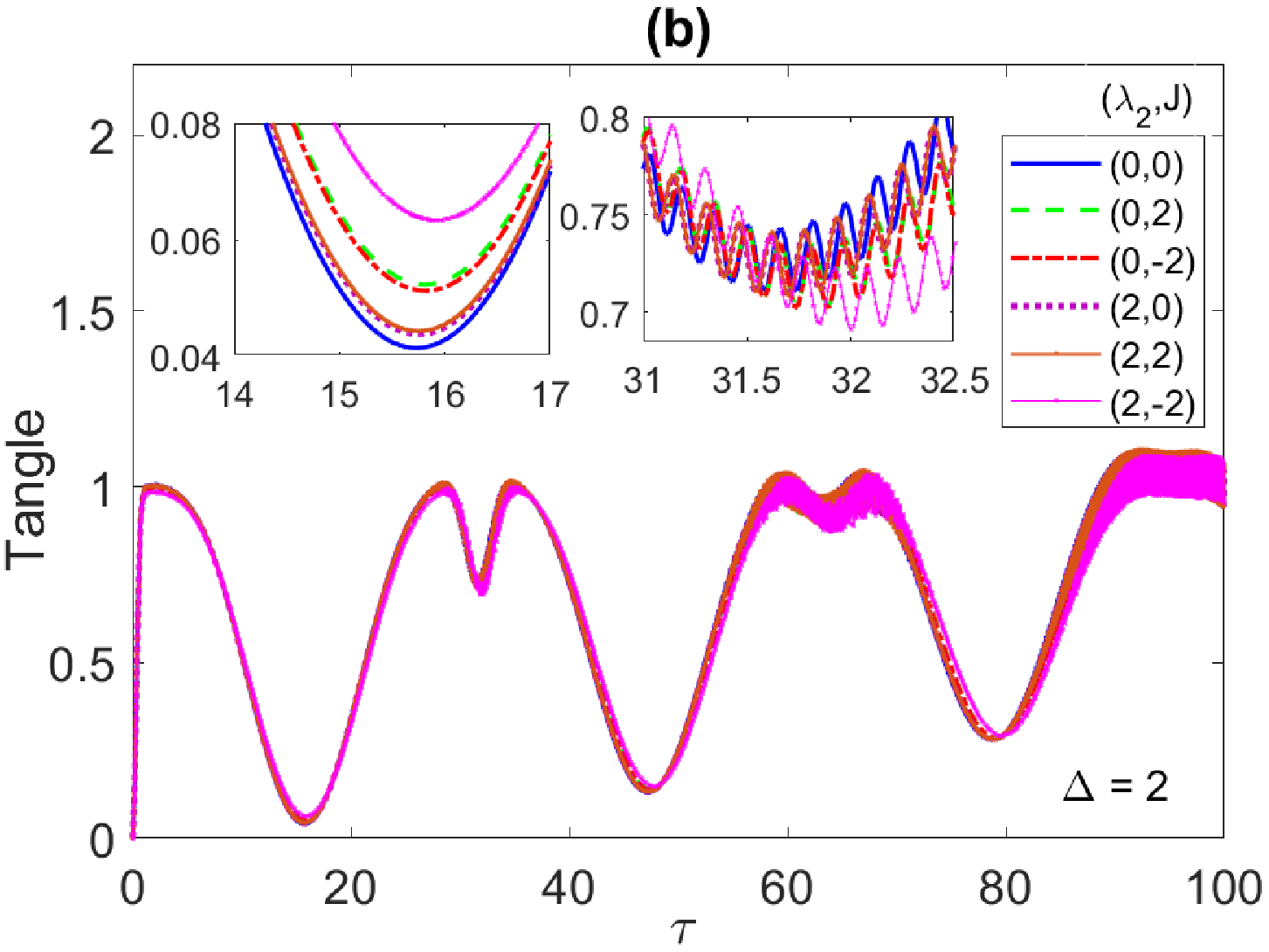}}\\
\subfigure{\includegraphics[width=6.4cm]{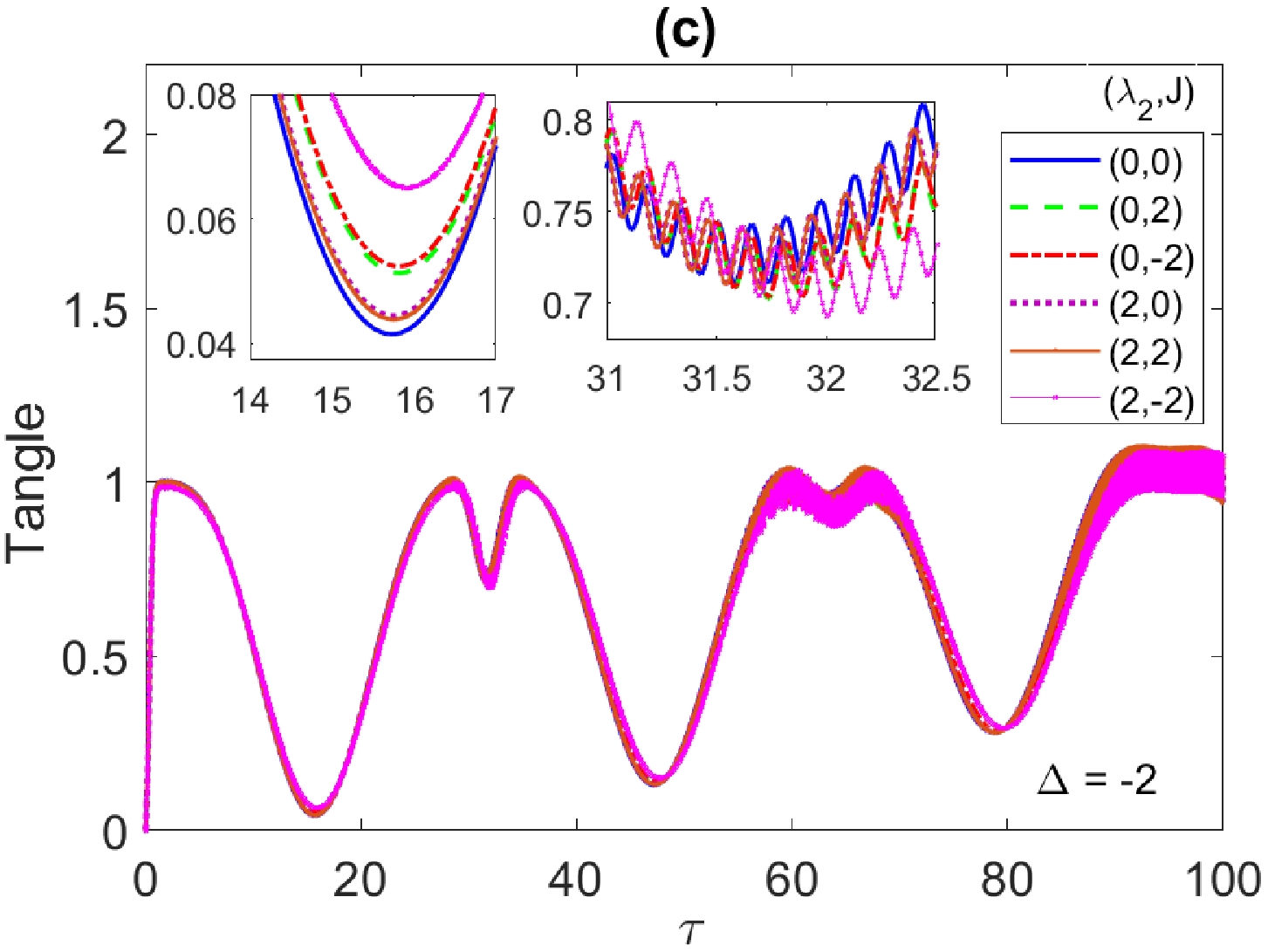}}\quad
\subfigure{\includegraphics[width=6.4cm]{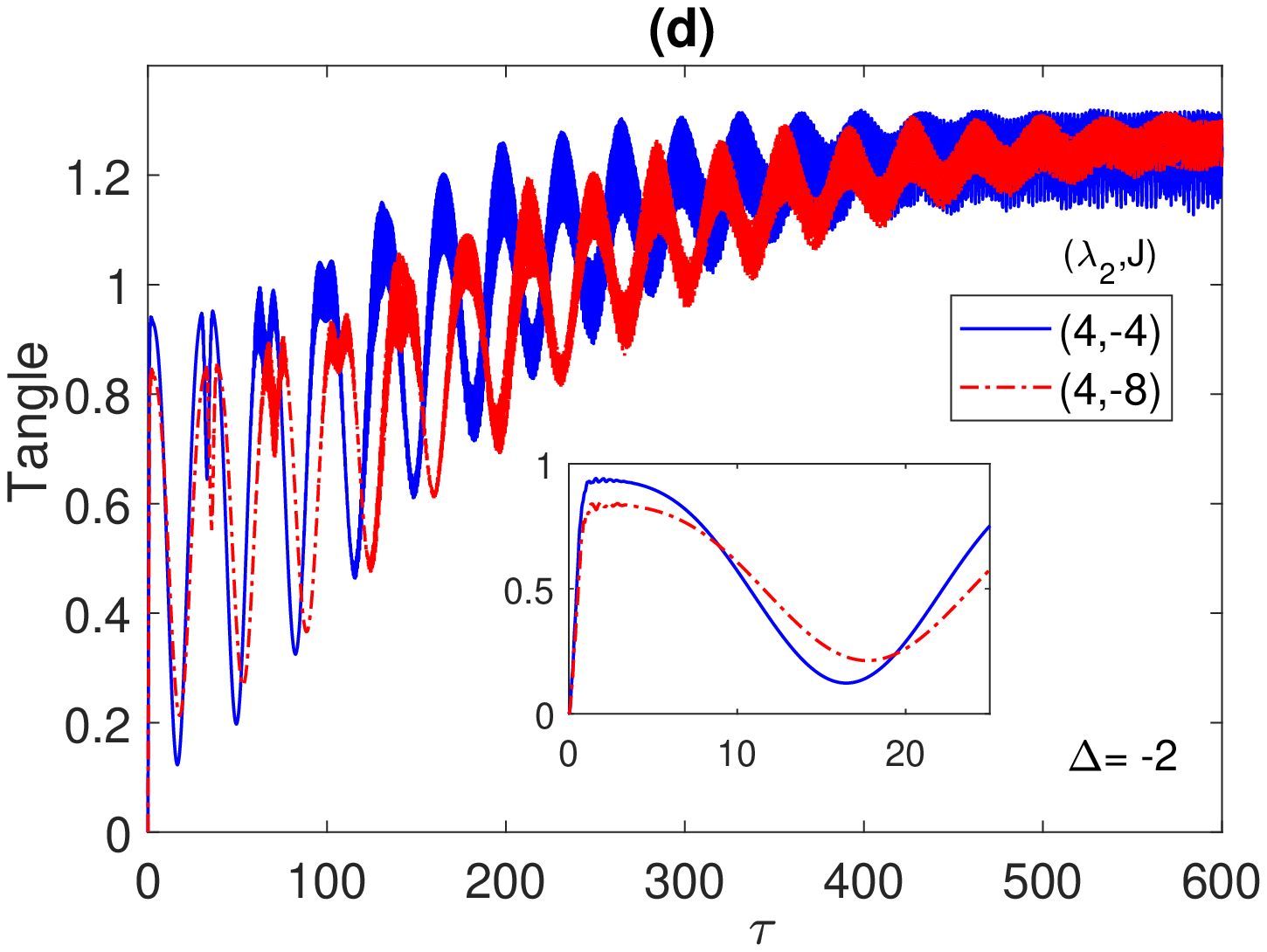}}\\
\caption{{
The tangle versus the scaled time $\tau=\lambda_1 t$ with the two atoms are initially  in a correlated Bell state $\psi_{Bc}=(\vert e_{1}\rangle \vert e_{2}\rangle+\vert g_{1}\rangle \vert g_{2}\rangle)/\sqrt{2}$ and the field is in a coherent state with $\bar{n}=100$, where in (a) $\Delta=0$; (b) $2$; (c) $-2$; (d) $-2$, while $(\lambda_2, J)$ take various values as shown in the legend in each panel.}}
\label{fig10}
\end{figure}

Clearly, comparing the results presented in the last three subsections shows that the dynamics of the system and its asymptotic behavior vary significantly depending on the initial state of the system. In general, when a single spin-1/2 particle is exposed to an external magnetic field, the magnetic dipole moment of the particle interacts with the field and precesses around its direction with a constant angle that depends on the initial state (orientation) of the spin with respect to the field. The precession (Larmor) frequency is determined by the magnetic field strength. On the other hand, when two spin-1/2 particles interact with each other, one of them could be considered as (a magnetic dipole) precessing in the field (magnetic moment) of the other one, where the dynamics of the system is determined by the relative orientation (initial state) of the spins, while the frequency by the interaction strength that depends on the anisotropy of the spin-spin coupling. This explains the critical impact of the initial state, as well as the relative strength of the coupling in the $x$, $y$ and $z$-directions, on the subsequent dynamics of the system and its asymptotic behavior.

\section{Quantum correlation between the two atoms and the radiation field}
\begin{figure}[htbp]
 \centering
\subfigure{\includegraphics[width=6.4cm]{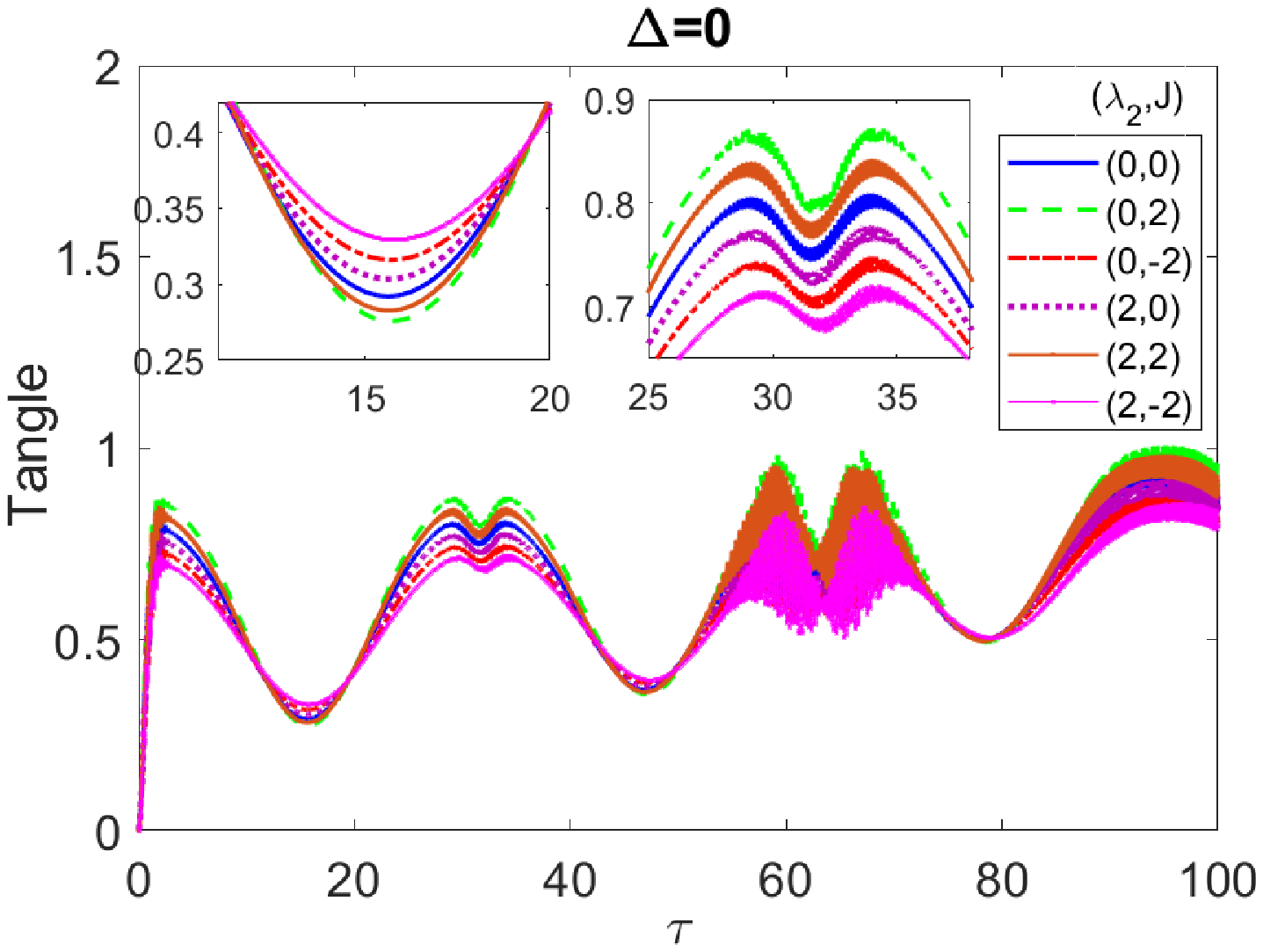}}\quad
\subfigure{\includegraphics[width=6.4cm]{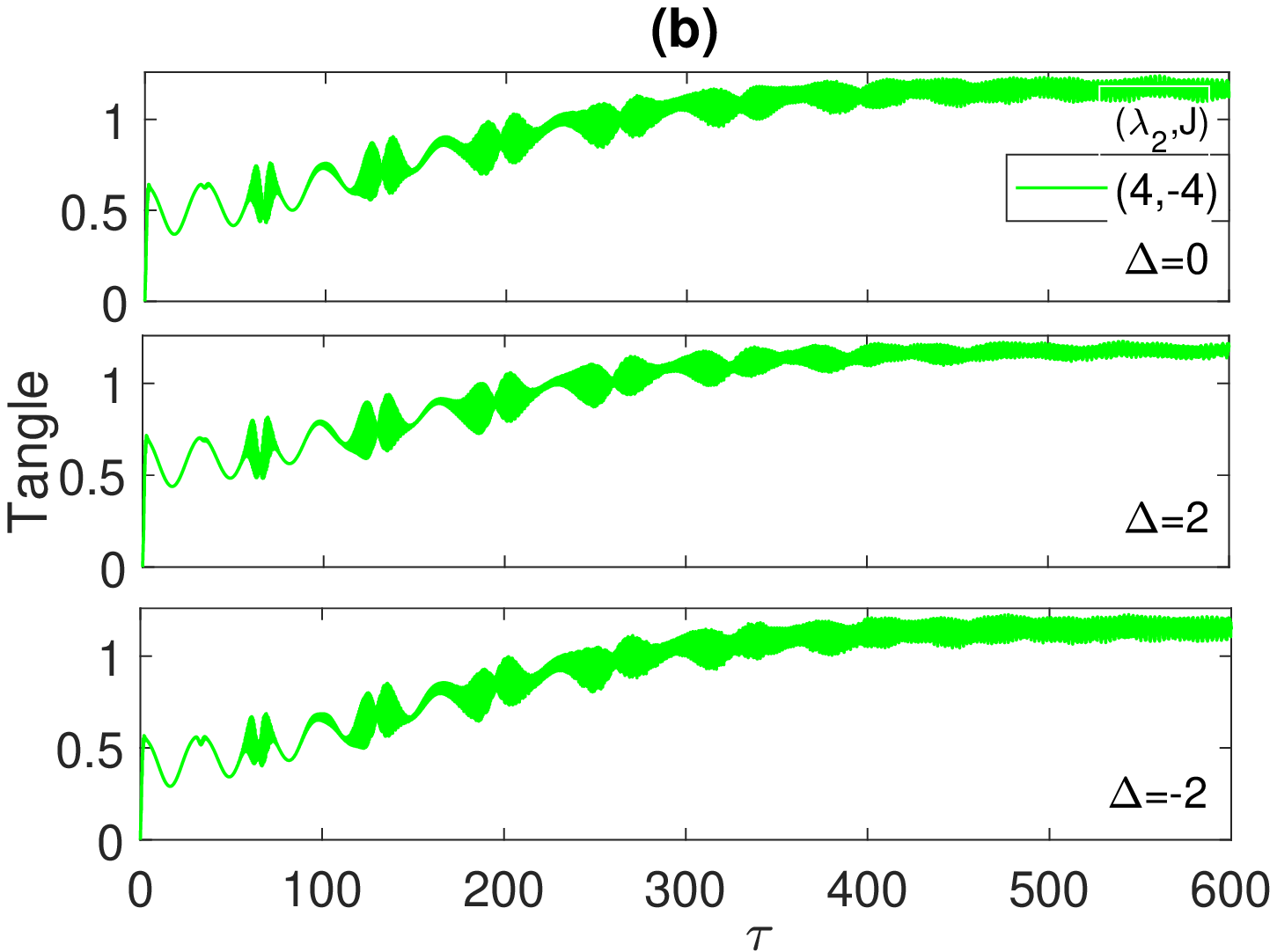}}\\
\subfigure{\includegraphics[width=6.4cm]{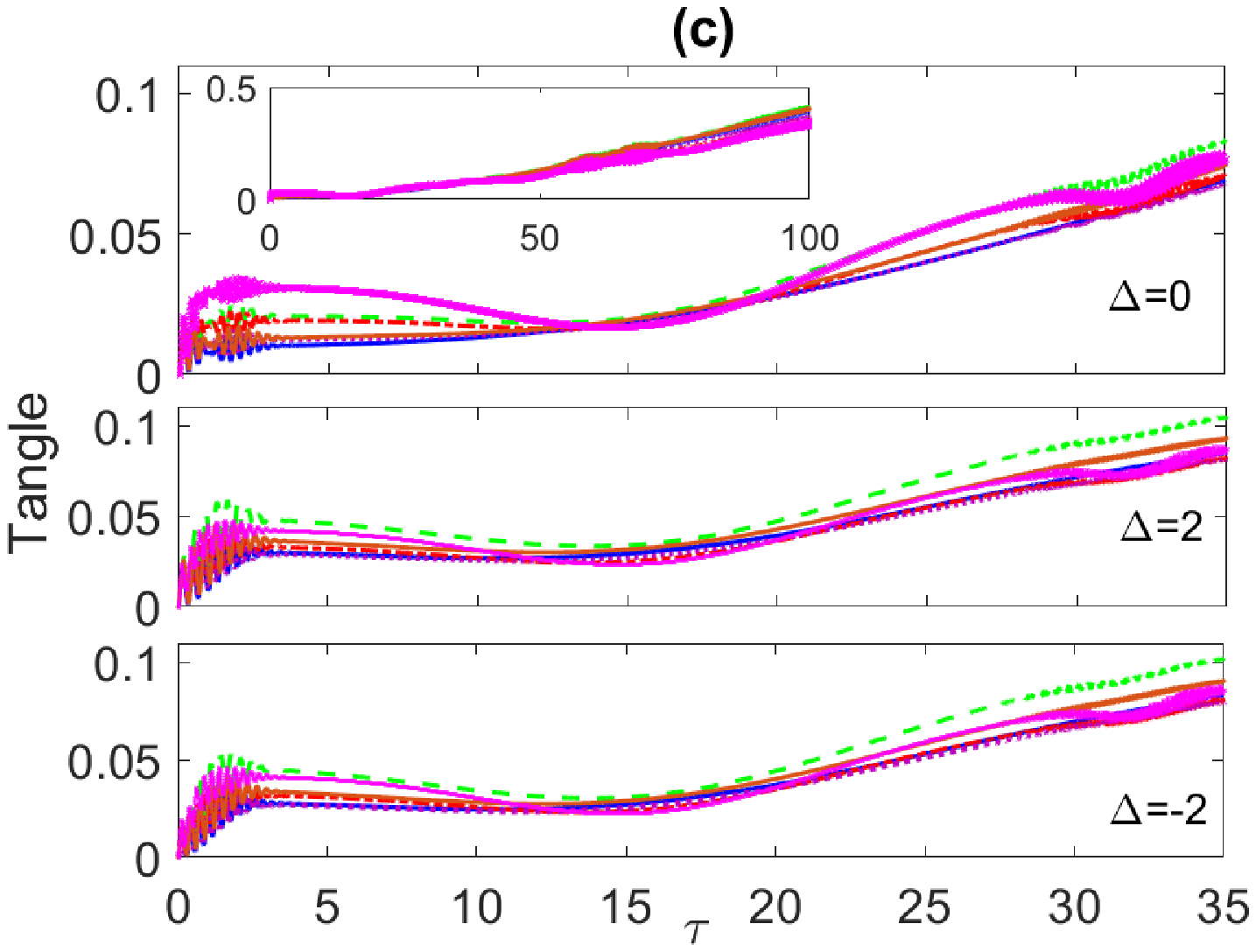}}\quad
\subfigure{\includegraphics[width=6.4cm]{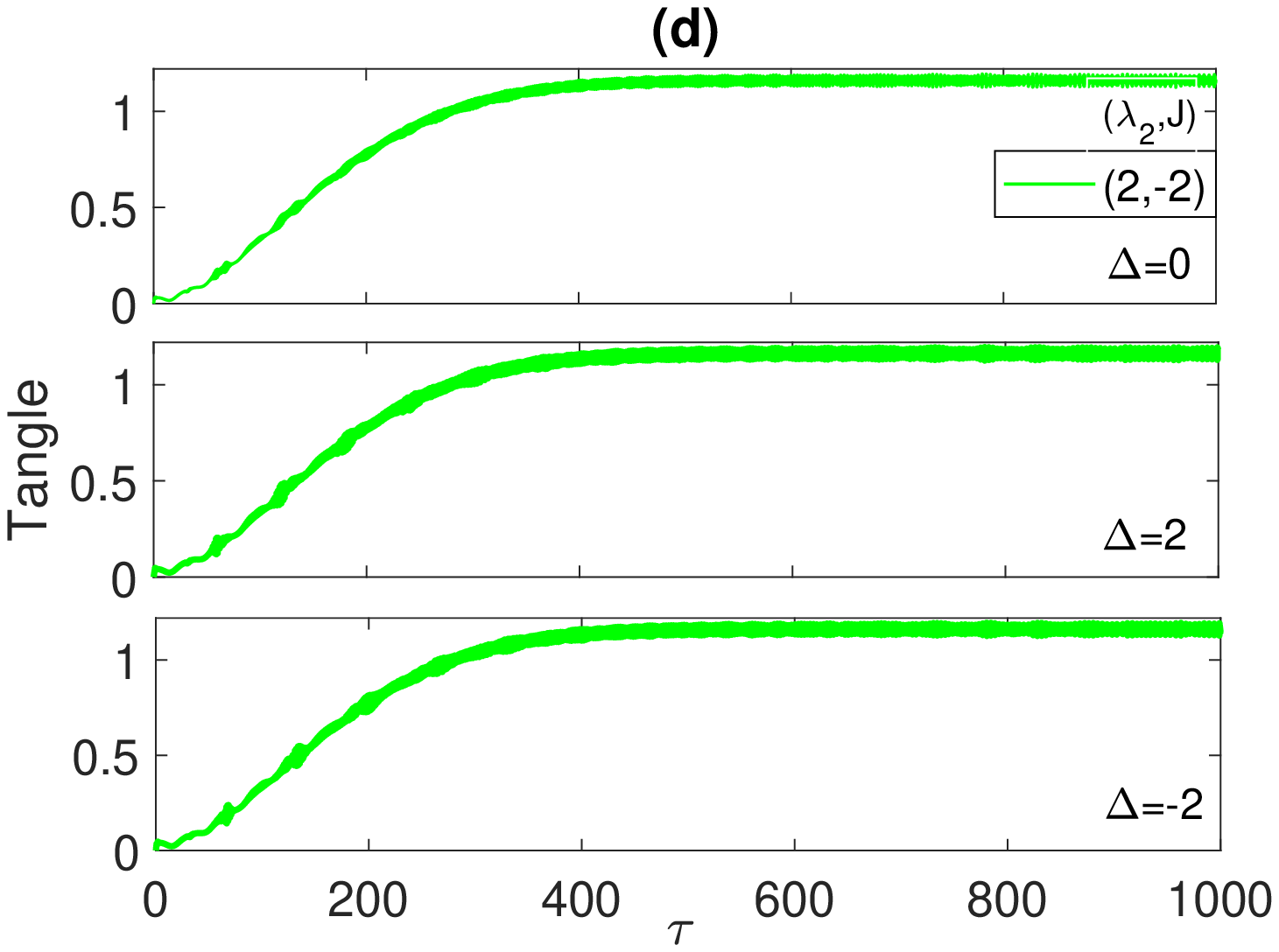}}\\
\caption{{
The tangle versus the scaled time $\tau=\lambda_1 t$ with the field is in a coherent state with $\bar{n}=100$ and the two atoms are initially in the W-like state $\psi_{W}$ in (a) and (b), and in the separable state $\psi_{L}$ in (c) and (d), at different values of the detuning as indicated in every panel. The values of the parameters $(\lambda_2, J)$ in panels (a) and (c) are as shown in the legend of Fig.~\ref{fig10}(a).}}
\label{fig11}
\end{figure}
It is very insightful to study the quantum correlation between the two atoms, $A_1$ and $A_2$, ensemble and the radiation field $F$, where the composite system is represented as $F \otimes\left(A_{1} A_{2}\right)$. This correlation can be utilized to shed some light on the reported behavior of the population inversion and entanglement dynamics that we have just discussed. Rungta, $\emph{et~al.}$, derived an analytic form for the concurrence of a bipartite system with arbitrary dimensions $d_A$ and $d_B$ in an overall pure state by generalizing the spin-flip operation to apply to higher dimensional systems \cite{Rungta:2001}. The resulting quantity, called the \emph{I-concurrence}, is given
by
\begin{equation}
C(\psi)=\sqrt{2 \nu_{A} \nu_{B}\left[1- Tr \left(\rho_{A}^{2}\right)\right]}
\end{equation}
where $\nu_A$ and $\nu_B$ are arbitrary scale factors that, in general, depend on the dimensions of the subsystems. The tangle $\tau$ of a bipartite system in a pure state with arbitrary subsystem dimensions reads
\begin{equation}
\tau_{A,B}(\psi) \equiv C^{2}(\psi)=2 \nu_{A} \nu_{B}\left[1- Tr \left(\rho_{A}^{2}\right)\right]
\end{equation}
In our case the tangle takes the form
\begin{equation}
\tau_{F\left(A_{1} A_{2}\right)}=2\left[1- Tr \left(\rho_{F}^{2}\right)\right]=2\left[1- Tr \left(\rho_{A_{1} A_{2}}^{2}\right)\right]
\end{equation}
In Fig.~\ref{fig10}, we depict the dynamics of the system tangle starting from the maximally entangled correlated Bell state. As can be noticed, at all the different detuning values $0, 2$ and $-2$, shown in panels (a), (b) and (c) respectively, the tangle starts at zero but rises abruptly to a value 1 within the same time period where the entanglement and the population have their initial rapid oscillations. Then the tangle decreases reaching a minimum in the middle of a time period that coincides with the population collapse and ESD periods. It peaks up again reaching a maximum value that is higher than the previous one, with a local minimum, within a time period that coincides with the entanglement and population revival oscillation periods. This behavior continues afterward as shown in the different panels, where the maxima and minima values increases with time and get closer while the time periods are shrinking in a similar fashion to the entanglement and population inversion profiles. The inner insets in panel (a) show that for $\Delta=0$ the curves coincide for $(0,\pm 2)$ as well as for (2,0) and (2,2), but they split at a non-zero detuning, as illustrated in the inner insets of panels (b) and (c). The asymptotic behavior of the tangle is presented in Fig.~\ref{fig10}(d), at $\Delta=-2$ at two choices of the parameters; (4,-4) (blue line) and (4,-8) (red line). The tangle profile is oscillatory with a decreasing amplitude and increasing mean value that reach an approximately constant value asymptotically for both cases, which also agrees with the observed behavior of the entanglement and the population inversion. In fact, the other choices of the parameter values, at the different detuning values, show a similar asymptotic behavior, which we did not present here. In general, the different panels of Fig.~\ref{fig10} demonstrate that the quantum correlation between the ensemble of the two atoms and the radiation field minimizes during the population collapse and the ESD periods and maximizes during their revivals, when the energy exchange between the atoms and the field takes place. The tangle dynamics starting from an initial W-state is illustrated in Fig.~\ref{fig11}(a) and (b), where it shows a smaller oscillation amplitude compared with the Bell state case, at all detuning values but similarly it evolves to a quasi steady state asymptotically with an almost constant mean value. The detuning variation is not showing a noticeable effect on the tangle behavior in this case. In Fig.~\ref{fig11}(c) and (d), the time evolution of the tangle starting from the separable initial state is explored. The tangle starts from zero as in the two previous cases but this time it shows a much weaker oscillatory variation with a smaller amplitude and a faster increasing mean value, which reaches a quasi-steady state asymptotically as well. The effect of the non-zero detuning is to shift the tangle values up and slightly split out the tangle lines at the different parameters choices from each other. 
\section{Conclusion}
We studied a system of two two-level atoms (qubits) interacting off-resonance with a single-mode radiation field. We considered the two atoms to be coupled to each other through dipole-dipole and Ferromagnetic (anti-Ferromagnetic) Ising interactions. We presented an exact analytic solution for the time evolution of the system that spans its entire parameter phase space starting from any initial state. We utilized the analytic solution to study the entanglement dynamics, between the two atoms, and its asymptotic behavior, particularly when the system starts from an initial state that leads to entanglement sudden death (ESD) upon evolution. 
The combination of the Ising and the dipole-dipole interaction was found to be powerful in manipulating the ESD compared with either one of them separately, especially at non-zero detuning. Their combined impact was found to vary significantly depending on the type of Ising interaction (Ferromagnetic or anti-Ferromagnetic) and the initial state of the system.
The atomic population inversion and the quantum correlation between the two atoms ensemble and the radiation field synchronized with the entanglement dynamics, where the ESD periods coincided with the collapse ones in the population inversion and the valleys in the quantum correlations, while the revival oscillations of entanglement and atomic population matched the peaks in the quantum correlation. This means the entanglement revives from death and the quantum correlation maximizes when the two atoms exchange energy with the field. The asymptotic behavior of the system showed crucial dependence on the initial state, where the ESD was found to be always removable by tuning the system parameters except in the case of a maximally entangled initial correlated Bell state, where ESD was found to be persistent in that case for any choice of the parameters values.
The entanglement, atomic population and quantum correlations were found to synchronize and reach asymptotically a quasi-steady dynamic states of continuous irregular oscillation with a limited amplitude and an approximately constant mean value, indicating a continuous exchange of energy between the two atoms ensemble and the field along with a strong correlation between the atoms and the field, and diminished ESD upon tuning the system parameters.
\section*{Acknowledgment}
This work was supported by University of Sharjah, Office of Vice Chancellor of Research, grant No. 2002143094-P.

\end{document}